\documentclass[14pt]{article}
\pdfoutput=1
\usepackage{amsmath,amssymb,amsfonts,amsthm,bm,bbm,cancel,wasysym}
\usepackage{epsfig,graphics,graphicx,epstopdf}
\usepackage{array,booktabs,colortbl,colordvi,multirow}
\usepackage{colordvi,color,xcolor}
\usepackage{hyperref}
\usepackage{rotating}

\usepackage{verbatim}
\usepackage{cite}
\usepackage{subfig}
\usepackage{setspace}
\usepackage{url}
\usepackage[percent]{overpic}
\usepackage{slashed}
\usepackage{authblk}
\usepackage{xspace}
\usepackage{fullpage}
\usepackage{hyperref}

\def\ie{{\it i.e.}}
\def\eg{{\it e.g.}}

\newcommand{\qslash}{\cancel{q}~}
\newcommand{\pslash}{\cancel{p}~}

\def\to{\rightarrow}

\newskip\zatskip \zatskip=0pt plus0pt minus0pt
\def\matth{\mathsurround=0pt}
\def\lsim{\mathrel{\mathpalette\atversim<}}
\def\gsim{\mathrel{\mathpalette\atversim>}}
\def\atversim#1#2{\lower0.7ex\vbox{\baselineskip\zatskip\lineskip\zatskip
  \lineskiplimit 0pt\ialign{$\matth#1\hfil##\hfil$\crcr#2\crcr\sim\crcr}}}

\begin{document}


\begin{flushright}
SLAC-PUB-17598\\
\today
\end{flushright}
\vspace*{5mm}

\renewcommand{\thefootnote}{\fnsymbol{footnote}}
\setcounter{footnote}{1}

\begin{center}

{\Large {\bf Dark Moments for the Standard Model ?}}\\

\vspace*{0.75cm}

{\bf Thomas G. Rizzo}~\footnote{rizzo@slac.stanford.edu}

\vspace{0.5cm}

{SLAC National Accelerator Laboratory}\ 
{2575 Sand Hill Rd., Menlo Park, CA, 94025 USA}

\end{center}
\vspace{.5cm}


\begin{abstract}
\noindent  
If dark matter (DM) interacts with the Standard Model (SM) via the kinetic mixing (KM) portal, it necessitates the existence of portal matter (PM) particles which carry both dark and 
SM quantum numbers that will appear in vacuum polarization-like loop graphs. In addition to the familiar $\sim e\epsilon Q$ strength, QED-like interaction for the dark photon (DP), in 
some setups different loop graphs of these PM states can also induce other coupling structures for the SM fermions that may come to dominate in at least some regions of parameter 
space regions and which can take the form of  `dark' moments, \eg, magnetic dipole-type interactions in the IR, associated with a large mass scale, $\Lambda$.  In this paper, motivated 
by a simple toy model, we perform a phenomenological investigation of a possible loop-induced dark magnetic dipole moment for SM fermions, in particular, for the electron. We show 
that at the phenomenological level such a scenario can not only be made compatible with existing experimental constraints for a significant range of correlated values for $\Lambda$ 
and the dark $U(1)_D$ gauge coupling, $g_D$, but can also lead to quantitatively different signatures once the DP is discovered. In this setup, assuming complex scalar DM to satisfy 
CMB constraints, parameter space regions where the DP decays invisibly are found to be somewhat preferred if PM mass limits from direct searches at the LHC and our toy model setup 
are all taken seriously. High precision searches for, or measurements of, the $e^+e^- \to \gamma +{\rm DP}$ process at Belle II are shown to provide some of the strongest future constraints 
on this scenario. 
\end{abstract}
\vspace{0.5cm}
\renewcommand{\thefootnote}{\arabic{footnote}}
\setcounter{footnote}{0}
\thispagestyle{empty}
\vfill
\newpage
\setcounter{page}{1}



\section{Introduction}

The existence of dark matter (DM) on many scales in the universe has been widely established -- but what is it's true nature and does it interact with the particles of the Standard Model via 
non-gravitational forces? It seems more than likely that some kind of new interaction(s) is(are) necessary for the DM to achieve the relic density that was measured 
by Planck\cite{Aghanim:2018eyx}. Traditional DM matter candidates, such as Weakly Interacting Massive Particles(WIMPs)\cite{Arcadi:2017kky,Roszkowski:2017nbc} and 
axions\cite{Kawasaki:2013ae,Graham:2015ouw,Irastorza:2018dyq}, though far from being excluded, continue to have their parameter spaces slowly eroded by the lack of 
any signals in direct or indirect detection experiments or at the LHC\cite{LHC,Aprile:2018dbl,Fermi-LAT:2016uux,Amole:2019fdf} and this has opened the door to an ever widening array 
of possible DM models covering huge ranges in both DM masses and interaction coupling strengths\cite{Alexander:2016aln,Battaglieri:2017aum,Bertone:2018krk}. These new models 
also can 
involve novel DM production/annihilation mechanisms in order to reproduce the observed relic abundance besides that of the familiar thermal relic\cite{Steigman:2015hda,Saikawa:2020swg}. 
A large class of these new interactions can be described as being through either renormalizable (dim=4) or non-renormalizable (dim $> 4$) `portals' via which the DM interacts with the 
SM via the existence of a new set of intermediary particles with a wide range of possible masses. 

Of the various renormalizable portal frameworks, the kinetic mixing (KM)/vector portal setup\cite{KM,vectorportal} has received much attention in the literature and can be realized in many 
forms of various complexity.  At a minimum, this scenario assumes the existence of a new $U(1)_D$ gauge group, with a corresponding coupling constant, $g_D$, whose associated 
gauge gauge field, $V$, is termed the `dark photon' (DP)\cite{Fabbrichesi:2020wbt,Graham:2021ggy}. The SM fields are assumed to be singlets under $U(1)_D$ and thus carry no 
dark charge, \ie, $Q_D=0$. Since $U(1)_D$ is usually considered to be broken, as it will be assumed here, the simplest (and most easily generalized) possibility is that this occurs via 
the vev(s) of at least one dark Higgs field in analogy with the spontaneous symmetry breaking in the SM. In particular, of interest to us here will the familiar case where both the DP 
and DM masses, $m_{V,DM}$, lie in a comparable range below roughly $\sim 1$ GeV so that the DM can be a variant of the traditional thermal relic albeit via the new non-SM interactions.

Within this kinetic mixing/vector portal scenario, there are only a few familiar ways for the dark photon to couple to the fields of the SM. 
In order for the KM portal itself to operate, new particles, here termed portal matter (PM)\cite{Rizzo:2018vlb,Rueter:2019wdf,Kim:2019oyh,Wojcik:2020wgm,Rueter:2020qhf},  must also 
exist which carry both SM as well as dark quantum numbers which then appear in vacuum polarization-like graphs connecting, \eg, the SM $U(1)_Y$ hypercharge gauge field with 
the corresponding field of the $U(1)_D$ DP. Assuming that the DP mass scale is $\lsim 1$ GeV, as we will below, after canonically normalizing all the fields to remove the KM and after 
both SM and $U(1)_D$ symmetry breaking occurs, this KM leads to a coupling of the DP to SM fields of the well-studied form $\simeq e\epsilon Q_{em}$. Here the dimensionless 
parameter $\epsilon$, usually considered for successful phenomenology to lie roughly in the $\sim 10^{-3}-10^{-4}$ range given the DM/DP mass region we are considering, is 
the strength of this KM which in UV-models can be analytically determined from the masses and quantum numbers of the PM fields participating in this loop graph, apart from an 
overall factor of $g_D$. In fact, for PM fields with O(1) mass splittings, one finds that values of $\epsilon$ in this interesting range are a natural expectation of this type of scenario and so these 
well-motivated values will be commonly assumed as part of the background to much of our discussion that follows below.

It is also well-known\cite{Davoudiasl:2012ag,Davoudiasl:2012ig,Davoudiasl:2013aya,Davoudiasl:2014kua} that if any of the dark Higgs 
fields happen to carry SM quantum numbers then their (necessarily, compared to that of the SM Higgs) small vevs, $\lsim $ a few GeV, will induce a further mass mixing between the DP and the 
SM $Z$ which is of order $\delta \sim m_V^2/m_Z^2 \simeq 10^{-4} \sim \epsilon$. This provides the DP with additional couplings to the SM fields which are completely analogous to those 
of the $Z$ but which are suppressed by the above mixing factor, $\delta$. This, when considered along with the $\epsilon$-induced terms, are the ones usually studied when considering 
the potential couplings of the DP to SM fields. One of the most basic and important lessons of the KM portal scenario is that DM, and maybe the entire dark sector, only talks to us, \ie, the 
SM, via loops of new particles carrying both types of quantum numbers. But can such PM particles also generate other types of interactions between the SM and the dark sector? 

In this paper we attempt a preliminary, first pass examination addressing just this question: are other types of couplings of the DP to the SM possible or even dominant, particularly, if 
for some reason both the usual $\epsilon$ and $\delta$ generated terms are absent or suppressed and if so how would the physics of the well-known KM scenario picture be altered?  
One may easily imagine a situation where the dark Higg(s) fields do not carry any SM quantum numbers (so that $\delta=0$ naturally) or that those that do have only suppressed 
vevs{\footnote {In what follows, we will assume that $\delta=0$ or, at the very least, that its effects are negligible in comparison to the couplings we will discuss.}} and where 
accidental degeneracies and/or cancellations between the various contributions occur rendering the value of $\epsilon$ much smaller than the typical range described above. What else could 
happen in such a situation to generate a suitably large interaction so that, \eg, the DM annihilation to SM fields via the DP can yield the observed relic density while also satisfying 
other experimental constraints? One possibility is to take the PM approach a bit more seriously, as we do here, and consider an augmented new (perhaps not quite dark) gauge structure 
corresponding to a group,  
$G$, under which the PM fields, here considered to be vector-like fermions, and (at least some of) the SM fermion fields occur in the same representation; many scenarios of this form 
are possible\cite{Rueter:2019wdf,Wojcik:2020wgm}. When this large group, $G$, is broken, presumably at some scale above $\sim$ a few TeV or higher, leaving only $U(1)_D$ remaining  
unbroken in the dark sector at lower energies, the SM fermions remain uncoupled to $U(1)_D$ even though they transformed non-trivially under $G$ itself. There then will be new 
massive gauge bosons, which themselves carry dark charges, coupling to both the SM and PM fermion fields (as well as to the DP itself), that will generate loop-induced DP 
interactions with the SM beyond the well-known $\simeq e\epsilon Q_{em}$ coupling generated by the effects of KM alone. 

Paralleling the more familiar and long-discussed analyses of the neutrino's loop-induced couplings to the SM 
photon\cite{Bernstein:1963qh,Vogel:1989iv,Okun:1986hi,Kayser:1982br,Shrock:1982sc,Giunti:2014ixa}, as well as similar studies of the possible electromagnetic couplings of fermionic 
DM\cite{Barger:2010gv,Alves:2017uls,Chu:2018qrm,Coffey:2020oir,Lambiase:2021xcj,Fortin:2011hv,Pospelov:2000bq,Fitzpatrick:2010br,Hambye:2021xvd} and motivated by a very 
simple (but only partially phenomenologically realistic) toy model, in this paper we will consider these loop-induced couplings of the DP to the $Q_D=0$ SM fermion fields. In the 
P- and CP-conserving limit, the leading term in the IR, as we will show and as should not be surprising,  is a dark magnetic dipole moment interaction which we will then study in a 
semi-model-independent manner, while keeping within the PM framework, making comparisons as we progress to the more familiar vector KM-type interaction. 
We will see that in most cases much of the phenomenology remains qualitatively familiar, though with a quantitatively different behavior, when the associated mass scale of the dark 
magnetic dipole operator lies in the 10's-100's of TeV range and the process under consideration only probes energy scales below $\sim 1$ GeV. Due to the energy scaling behavior of 
this operator, the main tree-level exception to this result, as we will show, is the associated DP production process, \eg, $e^+e^-\to V\gamma$ at BaBar and BELLE II, where the somewhat 
higher energy scales, $\sim 10$ GeV, are being probed. Note that although one can perform a more general study of the EFT parameter space of a scenario such as this, here we will 
approach this problem with a strong bias coming from our interest in and previous examinations of models of PM and their predictions for the DP interactions with the fields of the SM.

The outline of this paper is as follows: In Section 2, we provide an overview of the basic setup that we will follow in our analysis as well as the presentation of a toy 
model for the guidance and motivation of our study. In Section 3, we consider this setup at the phenomenological level as applied to several processes that are typically employed as probes 
of DM/DP physics, \eg, the evaluation of the DM relic density, DM direct detection via scattering off of electrons, production of DM/DP in electron-nucleus scattering at fixed target 
experiments, in the decay of neutral pions and, finally, as noted previously, $e^+e^-\to V\gamma$ at BaBar and Belle II. In each case we will compare and contrast the results we obtain 
in the dark magnetic dipole moment scenario with the familiar $\epsilon$-suppressed ones resulting from the usual KM scenario. A discussion of our results, some possible future avenues 
of investigation and our conclusions are then presented in Section 4.


\section{Basic Framework and a Simple Toy Model}

It is well-known that the coupling of a photon to a neutral massive Dirac fermion, such as a Dirac neutrino or fermionic dark matter itself,  can be decomposed into a small finite set of 
various (effective) interaction terms of increasing dimension. Since the SM fermions, here denoted collectively by $f$, are assumed to be neutral with respect to the dark $U(1)_D$ 
gauge symmetry, an analogous decomposition can be performed for their interactions with the dark photon, $V$. We may write this interaction for the case of on-shell fermions (assumed to be 
flavor diagonal for simplicity) in a well-known manner as
\begin{equation}
{\cal L} = \bar f(p')\Gamma^{\rm {on}}_\mu f(p) V^\mu (q)\,,  
\end{equation}
with $q=p-p'$ and where for these on-shell external SM fermions we may write (here loosely employing the notation of Ref.\cite{Chu:2018qrm})  
\begin{equation}
\Gamma^{\rm {on}}_\mu = e\epsilon Q_f \gamma_\mu+\frac{i\sigma_{\mu\nu} q^\nu}{\tilde \Lambda_f}\big[M(q^2)+iE(q^2)\gamma_5\big]+\frac{(q^2\gamma_\mu-q_\mu \qslash)}{\tilde \Lambda_f^2}\big[C(q^2)-A(q^2)\gamma_5 \big]\,,
\end{equation}
where we have also allowed for the existence of the usual term generated by the kinetic mixing proportional to $\epsilon$. Here $\tilde \Lambda_f$ is some large 
mass scale, presumably $\gsim$ a few TeV or more, which will in general depend upon the identity of $f$ and $M,E,C$ and $A$ represent (dark) magnetic dipole, electric dipole, charge 
and anapole moment form factors, respectively, which as noted are generally functions of $q^2$.  
We might imagine expanding these form factors around their $q^2=0$ values in a power series as $1+ c_{i,f} ~q^2/\tilde \Lambda_f^2 +...$ where $c_{i,f}$ are a set of constants. 
It is instructive to consider the sizes of the various terms in this expression recalling that for experimental/phenomenological reasons we usually assume that $\epsilon$ lies in the 
$\sim 10^{-3}-10^{-4}$ range as is common in PM models. Since we will be concerning ourselves with a sub-GeV DM and DP, in what follows we will mostly be restricting ourselves to 
processes wherein the typical mass scales will be $\sim$ 
a few GeV of less, \ie, values $<< \tilde \Lambda_f$. This being the case, we can set the constants $c_{i,f}\to 0$ and drop the two terms in the effective coupling proportional to 
$q^2/\tilde \Lambda_f^2 << 10^{-4}$, corresponding to the charge and anapole form factors, in comparison to the two dipole moment terms. If, for simplicity, we also assume CP-conservation, 
then we are left only with the additional dark magnetic dipole interaction term so that we can then re-write the equation above under these assumptions as approximately given by 
\begin{equation}
\Gamma^{\rm {on}}_\mu \simeq e\epsilon_{eff} Q_f\gamma_\mu+\frac{ig_D\sigma_{\mu\nu} q^\nu}{\Lambda_f}\,, 
\end{equation}
employing a slight change in both both notation and normalization, and explicitly including a factor of the $U(1)_D$ gauge coupling, $g_D$, in the dark dipole moment coupling.  It is the effect 
of this dark magnetic dipole term on the couplings of the SM fermions to the DP that we will investigate further below under the assumption that the KM vector coupling piece of this 
interaction takes on its usually expected value or is for some reason sometimes somewhat suppressed although both terms may be present simultaneously. 

\begin{figure}[htbp]
\centerline{\includegraphics[width=2.5in]{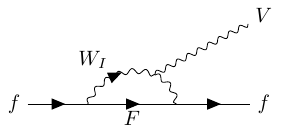}
\hspace{1.0cm}
\vspace {1.5cm}
\includegraphics[width=2.5in]{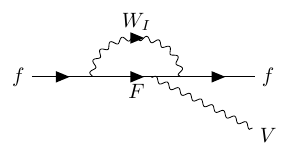}}
\vspace*{-0.20cm}
\caption{Typical 1-loop diagrams that can generate the dark magnetic dipole moment interaction in the toy model as discussed in the text. Note there are no graphs with dark photon, $V$, 
emission from the external lines as $f$ is a singlet under $U(1)_D$.}
\label{fig1}
\end{figure}

\begin{figure}[htbp]
\centerline{\includegraphics[width=5.0in,angle=0]{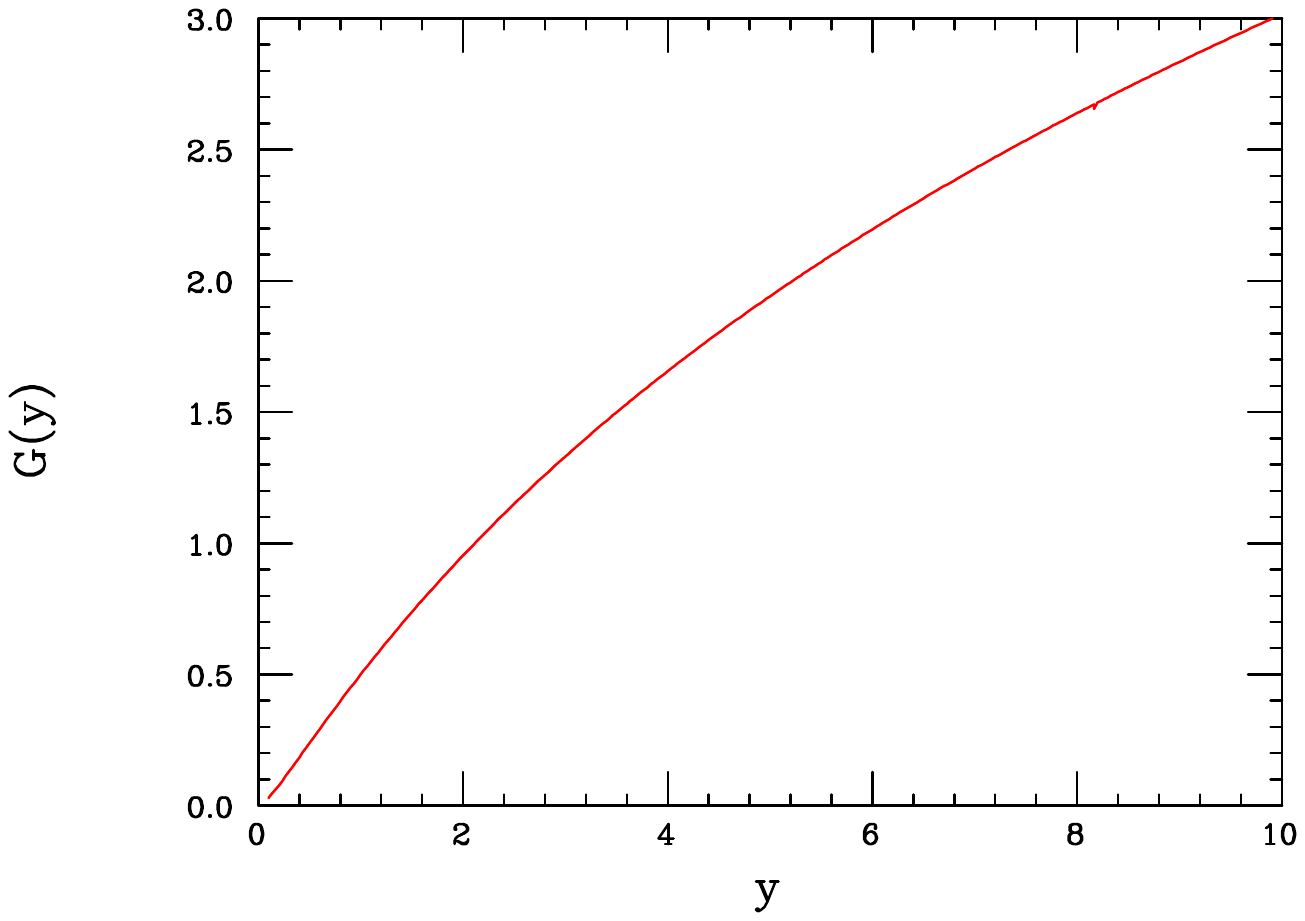}}
\vspace*{-1.50cm}
\caption{The function $G(y)$ as defined in the text for values of $y$ of O(1).} 
\label{fig2}
\end{figure}

Now we need to say a bit more about the magnitude that one might expect for the scale $\Lambda_f$. Clearly if the values of the $\Lambda_f$'s are too large this dark dipole term will have 
little phenomenological relevance. To motivate interesting scales we will employ a modified toy version of the dark sector model previously considered in 
Ref.\cite{Rueter:2019wdf} based on a SM-like, but fully broken, $SU(2)_I\times U(1)_{Y_I}$ dark gauge group. Here the corresponding gauge couplings are $g_I,g_I'$ in analogy with 
the SM and $g_D=g_I s_I$, $s_I$ here being the analog of the usual SM weak mixing angle, $s_W$. In this simple picture, (at least some of) the SM fermions, $f$, find themselves 
in $SU(2)_I$ doublet representations together with the vector-like fermion PM fields, $F$, carrying the same SM quantum numbers{\footnote {In the present discussion, for simplicity, we 
will ignore any potential mixing between the PM fields and those of the SM although these can play an important role in phenomenology.}}. The masses of the PM fermion fields as well as 
those of the gauge bosons in $SU(2)_I\times U(1)_{Y_I}$ apart from the DP, $V$, were all shown in earlier work to lie at or above the TeV scale to satisfy various LHC search constraints. The 
electrically neutral, but non-hermitian gauge boson within $SU(2)_I$, denoted as $W_I$, can then connect the SM field, $f$, with the corresponding doublet partner PM field, $F$, via a SM 
$W$-like interaction. Since both $F$ and $W_I$ carry $Q_D\neq 0$, they will couple to $V$ so that diagrams of the form similar to those shown in Fig.~\ref{fig1} will be generated at 
the 1-loop level. Assuming for simplicity that the $W_I$ couplings to these fermions are also vector-like, we can evaluate the contribution of these two diagrams to the effective dark magnetic 
dipole moment coupling in a straightforward manner by, \eg, employing the results as given in Ref.\cite{Leveille:1977rc}. Defining as usual $4\pi \alpha_D=g_D^2$, one finds that 
\begin{equation}
\frac{1}{\Lambda_f }=\frac{\alpha_D}{8\pi s_I^2 m_F}~G(y)\,, 
\end{equation}
where, if we imagine that $s_I^2\simeq s_W^2\simeq 0.23$, then numerically $8\pi s_I^2\simeq 5.8$. In this expression we have defined the mass squared ratio 
$y=m_F^2/m_{W_I}^2\sim$ O(1) so then the loop function $G(y)$ is given by 
\begin{equation}
G(y)=3y^2~\Big[\frac{-2(y-1)+(y+1)~ln(y)}{(y-1)^3}\Big]\,, 
\end{equation}
which is shown in Fig.~\ref{fig2}. Here we see that $G(y)$ is also generally an O(1) function so that, \eg, in the case of electrons, taking $m_E=1$ TeV, one could easily imagine that 
$\Lambda_e$ typically lies in the somewhat wide mass range of $\sim 5-1000$ TeV depending upon the values of both $s_I$ and $\alpha_D$, a mass range that we will return to 
below. Generically, one might imagine qualitatively similar parameter ranges for the other SM fermion flavors but with values scaling with their associated values of $m_F$. Similar sized 
contributions might also be expected to arise from any extended scalar sector in such a model as well. Note that in this toy model, since $SU(2)_I$ commutes with the SM gauge group, \eg, both 
the electron and the `electron neutrino' will have PM partners, $E$ and $N_e$, respectively, of essentially the same mass so that one might expect that $\Lambda_e=\Lambda_{\nu_e}$ 
since they originate from the same type of graphs with identical loop masses and gauge couplings. 

Of course, the PM fermions as well as these new gauge fields will be far too heavy to directly participate in any of the low energy processes we will be considering below. Although this is 
clearly only motivational, this simple model inspires us to consider a certain range of parameters when we analyze this situation further in the discussion below. As we 
will see, obtaining the observed relic density of DM will push us into a very similar mass range for the dark dipole coupling/mass scale. However, this present toy model is somewhat unrealistic 
in that, \eg, its limited PM and dark gauge sectors can also lead to too large of an anomalous (ordinary QED-like) magnetic dipole moment for the electron, $a_e$, which 
exceeds current experimental constraints and this obviously needs to be avoided in a realistic model. A more sophisticated PM particle spectrum with somewhat more complex gauge and dark 
Higgs interactions needs to be constructed to more successfully satisfy all of the various existing experimental requirements while simultaneously also yielding interesting values for the 
dark magnetic dipole moments.  The possibility of constructing such a more sophisticated dark 
sector model with the nice features above but without this $a_e$ drawback is a subject for future work and here we will generally treat the dark dipole moment interaction 
and the corresponding set of $\Lambda_f$ that we encounter below {\it only} as a simple phenomenological scenario with arbitrary parameters but with typical scales as motivated by 
this toy model, keeping this qualitative PM picture in mind.

As a side note, it is interesting to observe that these same type of PM fields in the toy model do {\it not} generate couplings of the DP to the SM $W,Z$ gauge bosons at 1-loop 
since these would appear only via a set of PM fermion triangle graphs. Since all the relevant couplings in such loops are assumed to be purely vector-like in such cases (in the absence 
of any mixing with the SM fields), charge conjugation remains a good symmetry in this sector of the model so that all such fermion loops will necessarily vanish. 

Since we expect the PM particles appearing in the loops that generate the dark magnetic dipole moments for the SM to have masses $\sim$ TeV, we will generally limit ourselves 
in our discussion below to processes whose energy scales are far below 1 TeV so that these loops are not resolved during the processes of interest (with one possible exception) 
and we can treat the dark magnetic dipole moment as a simple higher-dimensional interaction in an EFT-type framework.  Certainly, at LHC,  FCC-hh or even future $e^+e^-$ collider 
collision energies, this approximation that we will be making below can easily fail completely as the PM mass scales can be probed and the particles themselves may even 
be directly produced on-shell\cite{Rizzo:2018vlb,Rueter:2019wdf,Kim:2019oyh,Wojcik:2020wgm,Rueter:2020qhf}. However, we note in passing that our bias towards the PM model 
perspective of the physics underlying this EFT-type study colors our efforts to more fully explore the potentially much larger parameter space offered by a general EFT analysis, beyond 
the scope of the present work.

\section{Phenomenology}

\subsection{DM Relic Density}

The first, most obvious and important question that one might want to ask is: if the DP dominantly couples to the fermions of the SM through an effective dipole interaction, what range 
of $\Lambda_f$ values are needed to reproduce the observed relic density via a thermal mechanism assuming that the light DM annihilates via a virtual DP into SM fields? Recall 
that in our present discussion we are concerned with both DM/DP masses lying in the range below roughly $\sim 1$ GeV. Clearly if the required value of $\Lambda_f$ is too low, 
the scenario becomes immediately excluded since the PM fields cannot be too light thanks to LHC constraints. Before continuing, we remind the reader of two other familiar constraints 
with which we must abide when considering thermal DM in this mass regime: 
first, to avoid bounds from Big Bang Nucleosythesis, we must limit ourselves to DM masses in excess of roughly $\sim 10$ MeV\cite{Sabti:2019mhn}. Second, and perhaps 
more importantly, in this same DM mass range of interest to us here, roughly $\sim 10-1000$ MeV, data from the CMB (at $z\sim 1100$) constraints from Planck 
results\cite{Aghanim:2018eyx} inform us that during this period the DM annihilation cross section into various light SM charged states, \eg, $e^+e^-$, must be substantially 
suppressed\cite{Slatyer:2015jla,Liu:2016cnk,Leane:2018kjk,Bringmann:2006mu} in order to avoid the possible injection of any additional electromagnetic energy into the hot plasma. 
Recent analyses\cite{Cang:2020exa,Kawasaki:2021etm} of this constraint informs us that this bound lies roughly at the level of 
$\sim 5 \times 10^{-29}~(m_{DM}/100~ {\rm {MeV}})$ cm$^3$s$^{-1}$, 
noting that it depends approximately linearly on the DM mass. This is, in any case, roughly three orders of magnitude below that needed at freeze out to recover the observed relic 
density\cite{Steigman:2015hda,Saikawa:2020swg} of $\sim 4-8 \times 10^{-26}$ cm$^3$s$^{-1}$, with a more specific value depending upon the detailed nature of the DM. For example,  
if the DM is a complex scalar, as we will assume below, then for the range of masses of interest to us the target annihilation cross section is then 
$\simeq 7.5\sigma_0$\cite{Saikawa:2020swg} where 
$\sigma_0$ is defined as $\sigma_0=10^{-26}$ cm$^3$s$^{-1}$, the typical scale required for the thermal freeze-out mechanism. We also note in passing that 
there are additional constraints of a very similar magnitude for this range of DM masses from a completely different source which arise from 
Voyager 2 data\cite{Boudaud:2016mos,Boudaud:2018oya}. 

In what follows, in order to have a suppressed annihilation cross section during the CMB but still allowing for thermal freeze-out as noted earlier 
we will assume that the DM is a complex scalar, $\phi$, 
without a vev. This being the case, the annihilation through the single DP is a $p$-wave process and thus the thermally averaged annihilation rate, $<\sigma \beta_{rel}>$,  is proportional 
to the square of the relative velocities of the annihilating DM, \ie, $\sim \beta_{rel}^2$ which itself is proportional to the temperature, $T$. The temperature during the CMB era being much 
less than that during thermal freeze-out is then sufficient to provide the necessary cross section suppression factor. Using Eq.(3), we can then calculate the scalar DM annihilation 
cross section into a pair of SM fermions via an $s$-channel DP exchange, $\phi^\dagger \phi \to V^* \to \bar ff$, to be given by
\begin{equation}
\sigma_f =\frac{4\pi}{3}~\frac{N_cs\beta_\phi\beta_f}{(s-m_V^2)^2+(\Gamma_Vm_V)^2}~{\cal A}\,,
\end{equation}
where $\beta_{\phi,f}^2=1-4m_{\phi,f}^2/s$, $N_c$ is the standard color factor, $\Gamma_V$ is the DP total width and $s$ is the usual Mandelstam variable. Allowing for both interaction 
terms appearing in Eq.(3) to be simultaneously present, ${\cal A}$ is then given by the expression\cite{Atwood:1994vm}  
\begin{equation}
{\cal A}=Q_f^2\alpha \alpha_D \epsilon^2~\frac{(3-\beta_f^2)}{2}+\frac{1}{2} \Big(\frac{\alpha_D}{\Lambda_f}\Big)^2 s~(3-2\beta_f^2)+3\epsilon Q_f(\alpha \alpha_D s)^{1/2}~\Big(\frac{\alpha_D}{\Lambda_f}\Big) (1-\beta_f^2)^{1/2}\,,  
\end{equation}
where $Q_f$ is the fermion's electric charge. Here, the first term proportional to $\epsilon^2$ is the usual one from the vector coupling that we encounter in the KM/DP model while the 
second is that arising solely from the new dark magnetic dipole moment interaction we are considering. The final term in this expression is that which arises from the interference between the 
usual KM-induced  vector interaction proportional to $\epsilon$ and the new dark dipole moment piece and which we see vanishes far above threshold 
(the situation that we will consider here) as one might expect due to the different number of $\gamma$ matrices appearing in the two contributing matrix elements. 

Using the above cross section expression, one is able to calculate the value of $<\sigma \beta_{rel}>$ to check the relic density. To do this we will {\it assume} here, as well as for the 
rest of our DM annihilation analysis (except where noted), that the magnetic dark dipole term is generally by far 
the dominant one contributing to this process since we want to explore the nature of this interaction in isolation. This means that we will effectively be setting $\epsilon \to 0$ in the 
immediate discussion that follows except where it is noted for comparison and generalization purposes.  We note that within the PM framework, this is generally an unlikely situation 
unless there are accidental degeneracies in the PM mass spectrum or accidental strong cancellations occuring in the calculation of the vacuum polarization-like diagrams leading to $\epsilon$.
To move forward, as is usual, we need to know, \eg,  the DP's total width to evaluate the DP propagator factor above. When  
$r=2m_\phi/m_V <1$, the DP can decay directly to DM pairs as $\Gamma_V(DM)/m_V =\alpha_D \beta_r^3/12$ where $\beta_r^2=1-r^2$; this represents a sizable 
partial width if $\alpha_D$ is substantial $\gsim 0.01-0.1$, say. However, when $r>1$, only decays to the SM fields via the dark magnetic dipole interaction 
are possible (in the limit that $\epsilon \to 0$) for which, \eg, in the case of the $e^+e^-$ final state, the partial width is given by (keeping the electron mass, 
$m_e$, finite for demonstration purposes here but neglecting it in any numerical DM cross section calculations) 
\begin{equation}
\frac{\Gamma_V(e^+e^-)}{m_V}|_{dipole}=\frac{\alpha_Dm_V^2}{6\Lambda_e^2} ~\Big(1+\frac{8m_e^2}{m_V^2}\Big) ~\Big(1-\frac{4m_e^2}{m_V^2}\Big)\,,
\end{equation}
which we expect to be rather suppressed, $\lsim 10^{-10}$ or so, in analogy to what occurs when the usual $\epsilon^2$ term instead dominates here as in the pure KM 
picture{\footnote {Note that in the limit that $m_e^2/m_V^2 \to 0$ limit for finite $\epsilon$ one can simply add the individual KM and dark 
dipole contributions to obtain a total width for $V$ when both types of interactions are present simultaneously.}} 

Although we will be considering only the dark magnetic dipole moment contribution to the DM annihilation process, it is interesting to compare the annihilation cross section in this case with the 
familiar $\epsilon^2$ one arising from the ordinary vector coupling that we 
obtain from KM.  Away from any SM mass thresholds so that the interference term in $\cal A$ can be neglected and taking $s\simeq 4m_\phi^2$, which is a fair estimate during freeze-out since 
velocities are still relatively small, the ratio of the dark magnetic dipole moment and the standard KM DM annihilation cross sections is given simply by 
\begin{equation}
{\cal R} \simeq \frac{1}{2\alpha\alpha_D} \Big(\frac{2m_\phi\alpha_D}{\epsilon\Lambda_f}\Big)^2\,,
\end{equation}
so that if we assume that, \eg, $m_\phi=100$ MeV, a `traditional' value of $\epsilon=10^{-4}$ and $\Lambda_f/\alpha_D=25$ TeV for demonstration purposes only, as well as the existence of 
only a single SM final state, we find that 
${\cal R}\simeq 0.44/\alpha_D$, which, if then taken equal to unity, would imply that $\Lambda_f \simeq 11$ TeV, which is not an unexpected result given the discussion above. Of course 
in the analysis to follow we will generally assume that $\epsilon$ is far smaller so that the dark magnetic dipole term represents the by far dominant contribution except where noted for 
purposes of comparison.

Before performing a detailed numerical study it is interesting to examine some of the scaling features of the pure dark magnetic dipole piece of the cross section in the region below the 
resonance and make a comparison with the usual KM situation; this can be easily done by employing Eqs.(6) and (7) above. With $r$ as defined 
above and taking $\gamma=\sqrt s/2m_\phi$ (recalling that $\gamma$ never becomes too large in the thermal bath near freeze out), we can pull out overall factors of the DP mass, 
$m_V$, from the expressions above. Thus, \eg, in the expression for ${\cal A}$ in Eq.(7), we make the replacement $s=4m_\phi^2\gamma^2=(m_Vr\gamma)^2$ so that for $\gamma \simeq 1$, 
different powers of $r$ will appear in each of the 3 contributing terms appearing there. Correspondingly, the $s$ appearing in the numerator of the factor in front of ${\cal A}$ in Eq.(6) can 
be expressed in a similar manner leading to an overall factor of $r^2$.  Thus, in the usual $\epsilon^2$ contribution from the vector coupling, this results in a cross section with an overall 
familiar scaling of  $m_V^{-2}$ together with a factor of $r^2$ now in the numerator due to the single power $s$ 
appearing there. This implies that the familiar low-velocity cross section will grow reasonably quickly with $r$ until the effects of the DP resonance take over and that the cross section falls with 
increasing $m_V$, apart from the opening of new SM particle thresholds in the final state. For the pure dark magnetic dipole term, however, we see that {\it no} overall scale factors of $m_V$ will 
appear and that the numerator is found to instead scale as $r^4$ due to the additional power of $s$ now appearing there.  Not only will the cross section rise much faster with increasing 
$r$ (when $r <1$) but it is now be approximately $m_V$-{\it independent} apart from that which appears in the expression for DP's reduced total width, $\Gamma_V/m_V$, above as well as 
that due to the opening of various SM final state thresholds. Thus, for the case of the pure 
dark magnetic dipole operator term, the mass scaling of the DM annihilation cross section is essentially completely set only by the overall factor of $\Lambda_f^{-2}$.

\begin{figure}[htbp]
\centerline{\includegraphics[width=5.0in,angle=0]{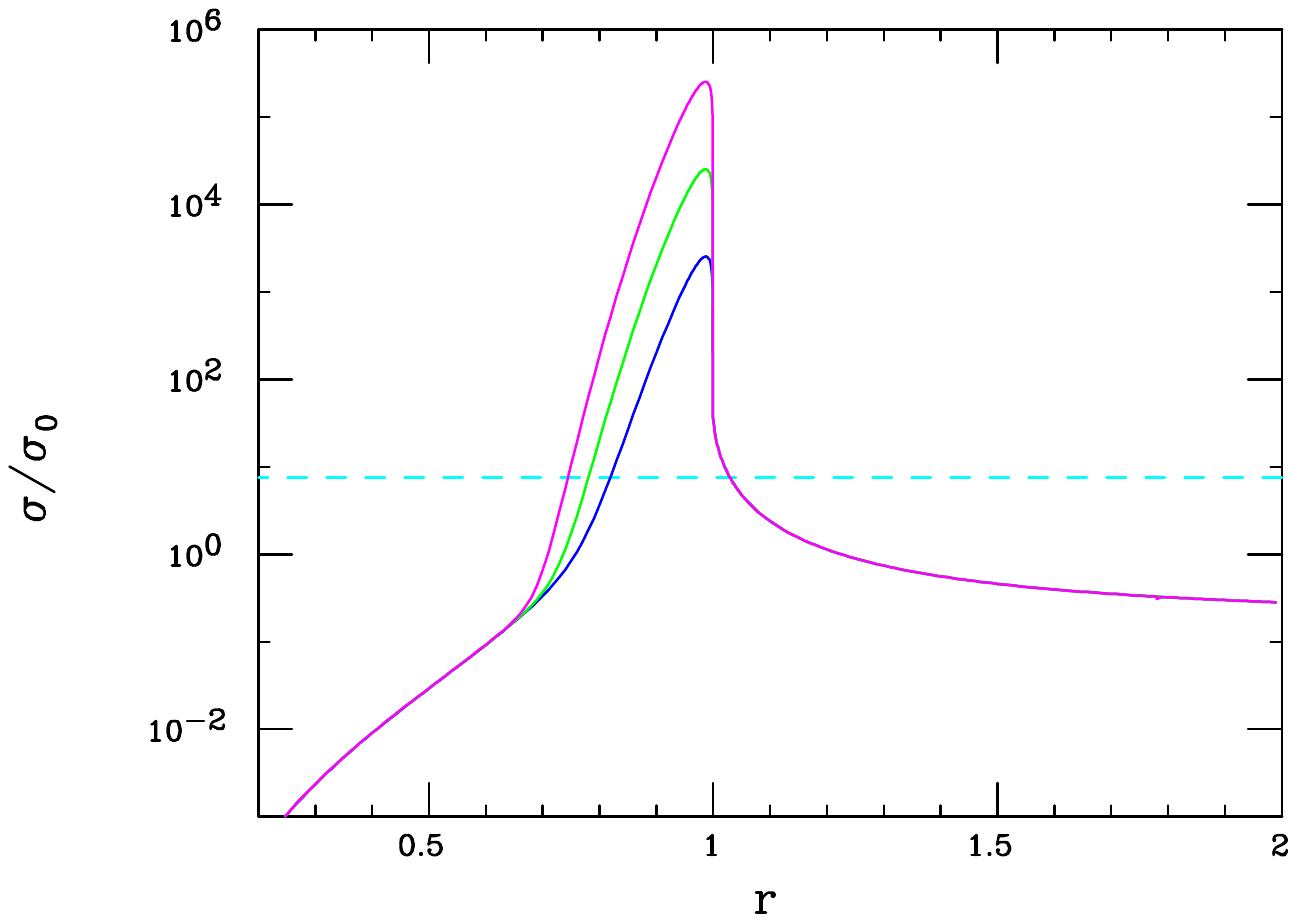}}
\vspace*{-2.3cm}
\centerline{\includegraphics[width=5.0in,angle=0]{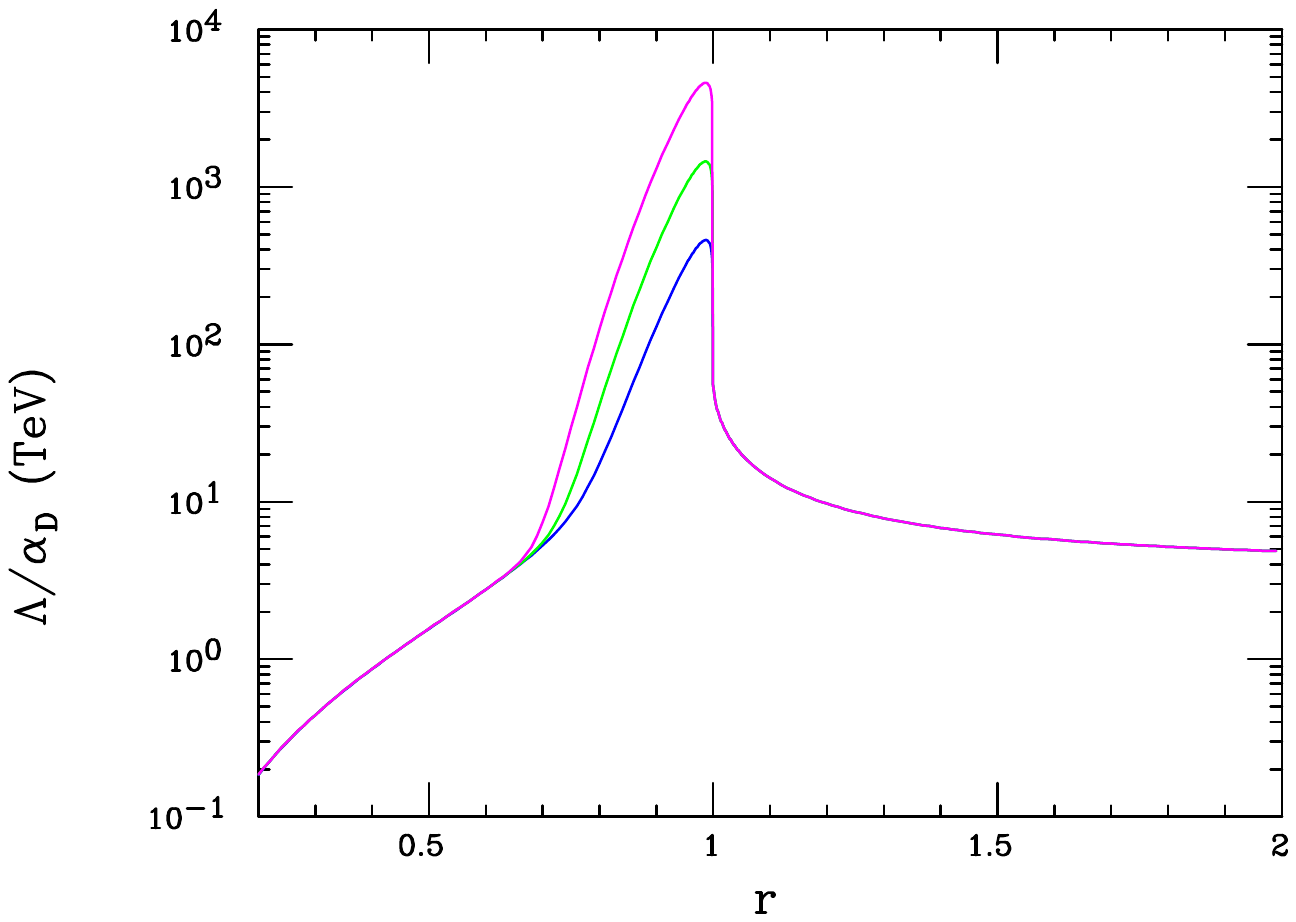}}
\vspace*{-1.30cm}
\caption{(Top) DM annihilation cross section into the $e^+e^-$ final state as a function of $r$ in units of $\sigma_0$ assuming that $\epsilon=0$, $x_F=20$, ${\cal F}=1$, and 
$\Lambda_e/\alpha_D=25$ TeV for purposes of demonstration. 
The blue (green, magenta) curve corresponds to choice $\alpha_D=1(0.1,0.01)$, respectively. The dashed cyan horizontal line corresponds to the cross section needed to reproduce the 
observed relic density. (Bottom) The value of $\Lambda_e/\alpha_D$ as a function of $r$ which reproduces the observed relic density again assuming that ${\cal F}=1$ as described in 
the text.}
\label{fig3}
\end{figure}

Turning now to DM annihilation, at freeze-out, we find after some algebra, that the thermal averaged cross section can be written as (see, \eg, Refs.\cite{Arcadi:2017kky,Plehn:2017fdg}) 
\begin{equation}
<\sigma \beta_{rel}>_{FO}=\frac{8x_F}{K_2^2(x_F)} ~\int_{\gamma_{min}}^\infty ~d\gamma~\gamma^2(\gamma^2-1)K_1(2\gamma x_F) ~\sigma_{\phi^\dagger \phi \to {\rm SM}}\,,
\end{equation}
where here, for demonstration purposes, the role of the `SM' is being solely played by the $e^+e^-$final state as above,  $x_F=m_\chi/T_{FO} \simeq 20$, $K_{1,2}$ are the standard 
modified Bessel functions and, as noted, $\gamma=\sqrt s/2m_\phi$ is the boost that the center of mass energy in the collision feels due to finite temperatures, with $\gamma_{min}=1$ here. 
Of course, in reality, multiple SM final states may be accessible as part of the DM annihilation process even if the DM is rather light. One reason for this when $\epsilon=0$ is that, at least in the 
toy model above, the induced dark magnetic dipole moments for the neutrinos and their leptonic partners are essentially identical. Thus, \eg, for 50 MeV DM, not only $e^+e^-$ but also 
at least one (or more) neutrino species may occur as part of the SM final state. To account for this, the actual cross section may need to be scaled upward by a factor 
${\cal F}=1+\sum_{i=1-3} (\Lambda_e/\Lambda_{\nu_i})^2$ which is equal to 2(4) if a single(all) neutrino species can appear as part of the final state with equal dark magnetic dipole scales. This 
does not happen in the usual KM setup but the general numerical impact is significant but not huge. 

The top panel in Fig.~\ref{fig3} shows the result for this annihilation rate calculation assuming that $\Lambda_e/\alpha_D=25$ TeV, $x_F=20$ and ${\cal F}=1$ with $\epsilon=0$ 
for purposes of demonstration. 
As expected, for low $r$ values, the rate rises quite rapidly with increasing $r$ until the strong influence of the resonance\cite{Feng:2017drg,Li:2015tka,Bernreuther:2020koj,Duch:2018ucs} 
appears which is clearly of some importance in reaching the desired cross section for the chosen set of input values and, as can be seen, somewhat depends upon the value of 
$\alpha_D$ via the size of the $V$ resonance total width. Note that when $r>1$, the required values of $\Lambda_e/\alpha_D$ tend to lie in the range $\sim 5-10$ TeV which can be 
in conflict with LHC constraints on the allowed PM mass range depending of course upon the other model parameters and how seriously we take our toy model.  
We note that large values of $r>2$ are excluded (at zero temperature) as in such a case the on-shell $2V$ production 
process, which is an $s$-wave, becomes possible thus violating the CMB constraints. In fact, even smaller values of $r$, above roughly $r \sim 1.6-1.7$ in the usual $\epsilon^2$ scenario, will 
become excluded by a similar argument once the $\phi^\dagger \phi \to VV^*,V^*\to e^+e^-$ process becomes sizable because it too is $s$-wave\cite{Rizzo:2020jsm}. In a parallel manner, 
we note that thermal effects can also excite the $2V$ final state on-shell as an $s$-wave even when $r$ is substantially below $r=2$ if $\alpha_D$ is large, as is well-known from studies of 
Forbidden DM Models \cite{Griest:1990kh,DAgnolo:2015ujb,Cline:2017tka,Fitzpatrick:2020vba,1837855,Rizzo:2021pxo}, so that the $s$-channel DP exchange process no longer 
dominates DM annihilation. Taken together all of these considerations somewhat disfavor the $r>1$ region to some extent in the present setup especially when $\alpha_D$ is large as it 
may be here. Thus in what follows we will generally assume that in all likelihood the DP decays invisibly, \ie, that the region $r<1$ is favored.

\begin{figure}[htbp]
\centerline{\includegraphics[width=5.0in,angle=0]{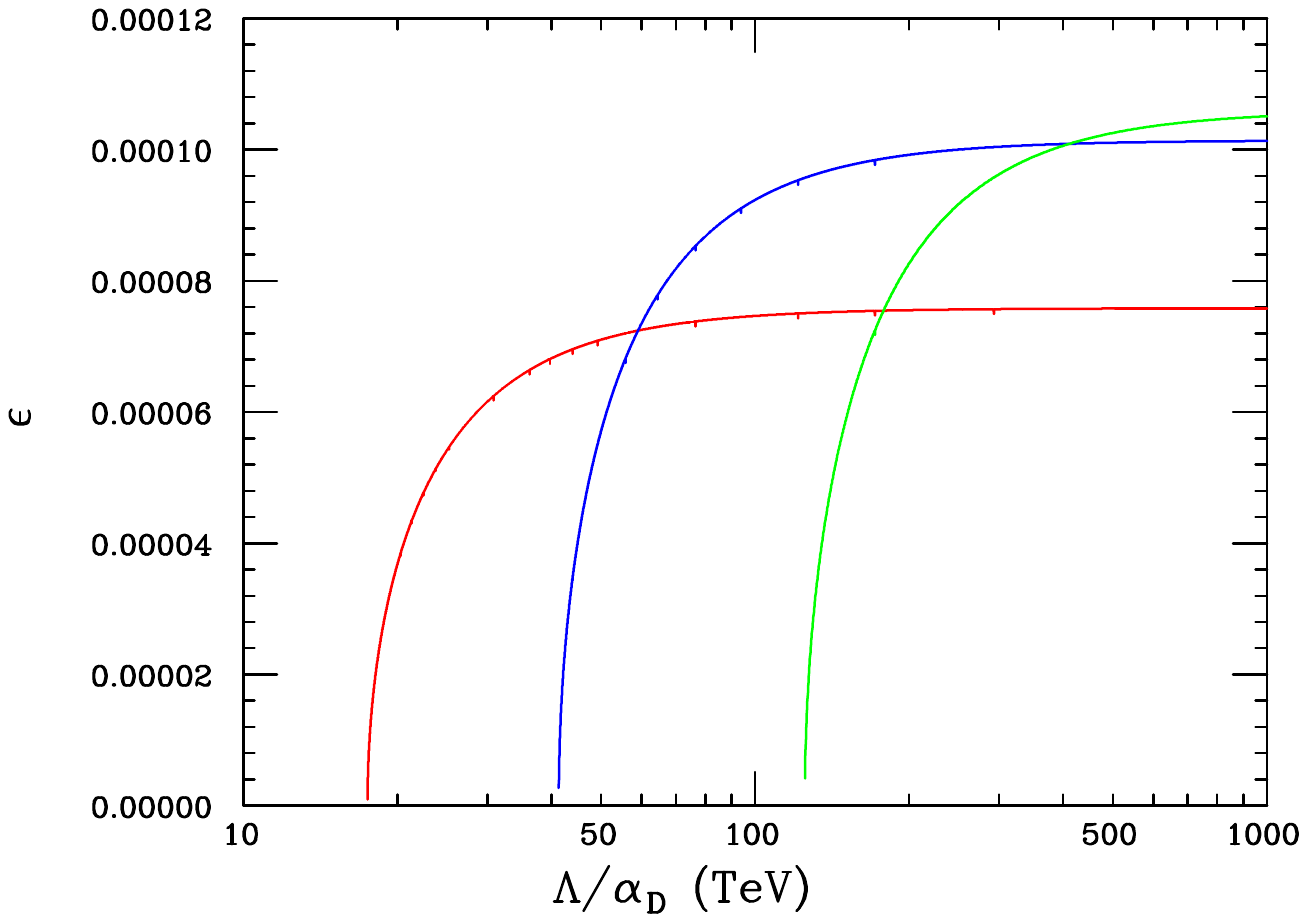}}
\vspace*{-1.50cm}
\caption{Correlated values of $\epsilon$ and $\Lambda_e/\alpha_D$ assuming both KM and the dark dipole magnetic moment contribute to the DM annihilation cross section yielding the 
observed relic density assuming that $r=0.8$ and $m_V=200$ MeV for purposes of demonstration. The red(blue, green) curves assume that $\alpha_D=1(0.1,0.01)$, respectively.} 
\label{fig10}
\end{figure}

Turning this calculation around, we can ask, as a function of $r$, what is the value of $\Lambda_e/\alpha_D$ that is needed to obtain the observed relic density; the result of this calculation 
is shown in the lower panel of Fig.~\ref{fig3} for the same assumed input values, $x_F=20$ and ${\cal F}=1$ as above. Several observations can immediately be made: ($i$) The value 
of $\Lambda_e/\alpha_D$ is, as expected, quite sensitive to the value of $r$ and, to a reasonable extent, $\alpha_D$, in the resonance region. ($ii$) Values of $r\lsim 0.7$ are 
somewhat disfavored as they require a dark dipole scale that is seemingly far too low and so likely conflicting with other, \eg, LHC constraints (as we have noted previously above). ($iii$) 
As the resonance is approached, acceptable values of $\Lambda_e/\alpha_D$ lying roughly in the rather wide range, $\sim 10-1000$ TeV,  for $r$ in 
the corresponding range $\simeq 0.7-0.9$ are obtained. ($iv$) Due to the rapid fall off observed above the $V$ resonance, (and given the arguments above) the region below the 
resonance is somewhat preferred, this being the regime where the DP can rapidly decay invisibly to DM as previously discussed. ($v$) Note that when ${\cal F}$ is larger than unity, as 
might be assumed in a realistic model, these preferred values of $\Lambda_e/\alpha_D$ that we obtain will be increased by a factor of $\sqrt {\cal F} \leq 2$ or so, depending 
somewhat weakly upon the DM mass as well as the relative dark magnetic dipole moment scales of the various contributing fermions.  
($vi$) Lastly, we note that it almost goes without saying that if the usual $\epsilon^2$ KM contribution to the cross section were also to be {\it simultaneously} significant then the corresponding 
values of $\Lambda_e/\alpha_D$ required to obtain the observed DM relic density could be easily significantly larger. The converse of this is also true: a finite value for $\Lambda_e/\alpha_D$ 
allows for a smaller value of $\epsilon$ needed to reproduce the observed relic density. In fact, for fixed values of $m_V$, $\alpha_D$ and $r$, a unique curve would then 
exist in the $\epsilon-\Lambda_e/\alpha_D$ plane which yields the correct relic density. As an example of this, we see for the rather typical input values of $r=0.8$ and $m_V=200$ MeV 
shown in Fig.~\ref{fig10}, an example of just this correlation. Here one observes that once $\epsilon$ becomes significant $\Lambda_e/\alpha_D$ easily gets pushed to larger values as is to be 
expected.

It is clear from this analysis that obtaining the necessary annihilation rate to reproduce the observed DM relic density is rather straightforward for a wide range of dark magnetic dipole 
moment scales but for only a correspondingly rather restricted range in ratio of the dark matter to dark photon masses under the assumption that the DM is a complex scalar.

\subsection{Direct Detection}

As a further test of this scenario, we next consider the capabilities of existing and future planned direct detection experiments to probe parts of the dark magnetic dipole moment 
model parameter space which yield the observed DM relic density. Due to the relevant DM/DP mass ranges that we consider here, both below $\sim 1$ GeV, we will restrict our attention 
to the elastic scattering of complex scalar DM with a free electron initially at rest as our benchmark as this is likely the most representative process for this mass range as 
is well-known\cite{Aguilar-Arevalo:2019wdi,Aprile:2019xxb,Aprile:2020tmw,Amaral:2020ryn,Arnaud:2020svb,Barak:2020fql}.  Canonically, for a bound electron, the relevant scattering cross 
section is expressed 
in the form $\sigma_{e\phi}=\bar {\sigma_e} {\cal G}^2$, where $\bar {\sigma_e}$ is the reference cross section for the scattering of DM off of a free electron at rest due to KM and the form 
factor ${\cal G}$ parameterizes the detailed nature of the target electron including, \eg, binding effects{\footnote{See, for example, the work in 
Refs.\cite{Knapen:2021run,Griffin:2021znd,Chen:2021qao,Baxter:2019pnz,Essig:2019xkx}} and can be both DM velocity and momentum transfer-dependent. 

In our calculation will assume that $m^2_e<< m^2_{DM,V}$, which we 
expect to be a good approximation based on BBN constraints\cite{Sabti:2019mhn}, $m_\phi=m_{DM}\gsim 10$ MeV, and, since we are here specifically interested in how the dark magnetic 
dipole moment influences the value of the reference cross section $\bar {\sigma_e}$ {\it itself},  we will further assume that the initial electron is both free and 
at rest in our analysis taking ${\cal G}=1$ and thus neglecting any  
structure effects arising from the bound state nature of the electron in the atom and the possible effects of the atom's binding in whatever medium it may lie. We note that 
even before we begin our calculation we expect that the pure dark magnetic dipole moment induced scattering piece of the total cross cross section, $\sigma_{e\phi}$, will be quite suppressed 
in comparison to the usual KM result from the vector coupling proportional to $\epsilon^2$ due to the small value of the momentum transfer ($q^2=-Q^2$) direct dependence of the dark 
dipole interaction above arising from the necessarily small DM velocity in the galaxy, $\beta_\phi \simeq 10^{-3}$. As is also standard, we will make the approximation that $Q^2<<m_V^2$ 
in the DP propagator and note that 
in this non-relativistic limit, given the previously mentioned approximations, $s\simeq m_\phi^2$. With this, even when the value of $\epsilon$ is for some reason very suppressed, it is important to 
consider all of the cross section terms arising from Eq.(3) as we will see below.

Given Eq.(3), we can directly determine the desired $e-\phi$ elastic scattering cross section which, after integrating over $Q^2$ and making the approximations above, is 
given by ($Q_e^2=1$) 
\begin{equation}
\sigma_{e\phi}\simeq \frac{16\pi m_e^2}{m_V^4}\alpha \alpha_D \epsilon^2 ~\Big(1+2Q_e \chi +\frac{10}{3}\chi^2\Big)\,,
\end{equation}
where we have factored outside the usual $\epsilon^2$ piece arising from KM vector coupling and $\chi$ is then just the dimensionless ratio 
\begin{equation}
\chi=\frac{1}{(\alpha \alpha_D \epsilon^2)^{1/2}}~ \Big(\frac{\alpha_D}{\Lambda_e}\Big)~m_e\beta_\phi^2\,,
\end{equation}
and essentially measures the relative strength of the dark magnet dipole moment interaction in comparison to that of the usual KM $\epsilon$ term which will be comparable when $\chi \sim O(1)$. 
Here we see that the pre-factor of this expression yields the canonical reference cross section arising from KM alone, $\bar{\sigma_e}$.  
Employing the typical parameter values from the previous discussion above we find, however, that not unexpectedly $\chi <<1$: 
\begin{equation}
\chi \simeq \frac{2.4\cdot 10^{-9}}{\sqrt{\alpha_D}}~ \Big(\frac{10^{-4}}{\epsilon}\Big)~\Big(25~ {\rm TeV} ~\cdot\frac{\alpha_D}{\Lambda_e}\Big)~\Big(\frac{\beta_\phi}{10^{-3}}\Big)^2\,,
\end{equation}
from which we see that $\epsilon$ would need to be very highly suppressed before the dark dipole term becomes dominant or even relevant, assuming both couplings are simultaneously 
contributing to this process. 
For example, in Fig.~\ref{fig11}, we see how turning on a finite value for $\epsilon$ rapidly overcomes the dark magnetic dipole moment contribution to this scattering cross section. For (very) 
small values of $\epsilon$ only the dark dipole piece contributes so that the cross section is essentially $\epsilon$-independent. Then as $\epsilon$ grows there is first some initial destructive 
interference between the two contributions (since $Q_e=-1$) and then as $\epsilon$ grow further, to $\sim 10^{-12}$ and beyond,  the KM piece fully dominates and the cross section shows 
the usual $\epsilon^2$ growth. For completeness, we note that if the limit $\epsilon \to 0$ were to be actually realized to a good approximation, one finds in this case that the pure 
dark magnetic dipole cross section to be just
\begin{equation}
\sigma_{e\phi}(\epsilon\to 0)\simeq \frac{160\pi m_e^4}{3m_V^4}~\Big(\frac{\alpha_D}{\Lambda_e}\Big)^2\beta_\phi^4\,,
\end{equation}
or, numerically, using the suggestive input values from above, yielding  
\begin{equation}
\sigma_{e\phi}(\epsilon\to 0)\simeq (7.1\cdot 10^{-56} ~{\rm cm^2})~\Big(25~ {\rm TeV} ~\cdot\frac{\alpha_D}{\Lambda_e}\Big)^2~\Big(\frac{100 ~{\rm MeV}}{m_V}\Big)^4~\Big(\frac{\beta_\phi}{10^{-3}}\Big)^4\,,
\end{equation}
which is essentially at an invisible level, far below the neutrino floor, for the foreseeable future as we might have expected. Thus in the pure dark magnetic dipole interaction limit, 
direct detection experiments should see no DM signal. 

\begin{figure}[htbp]
\centerline{\includegraphics[width=5.0in,angle=0]{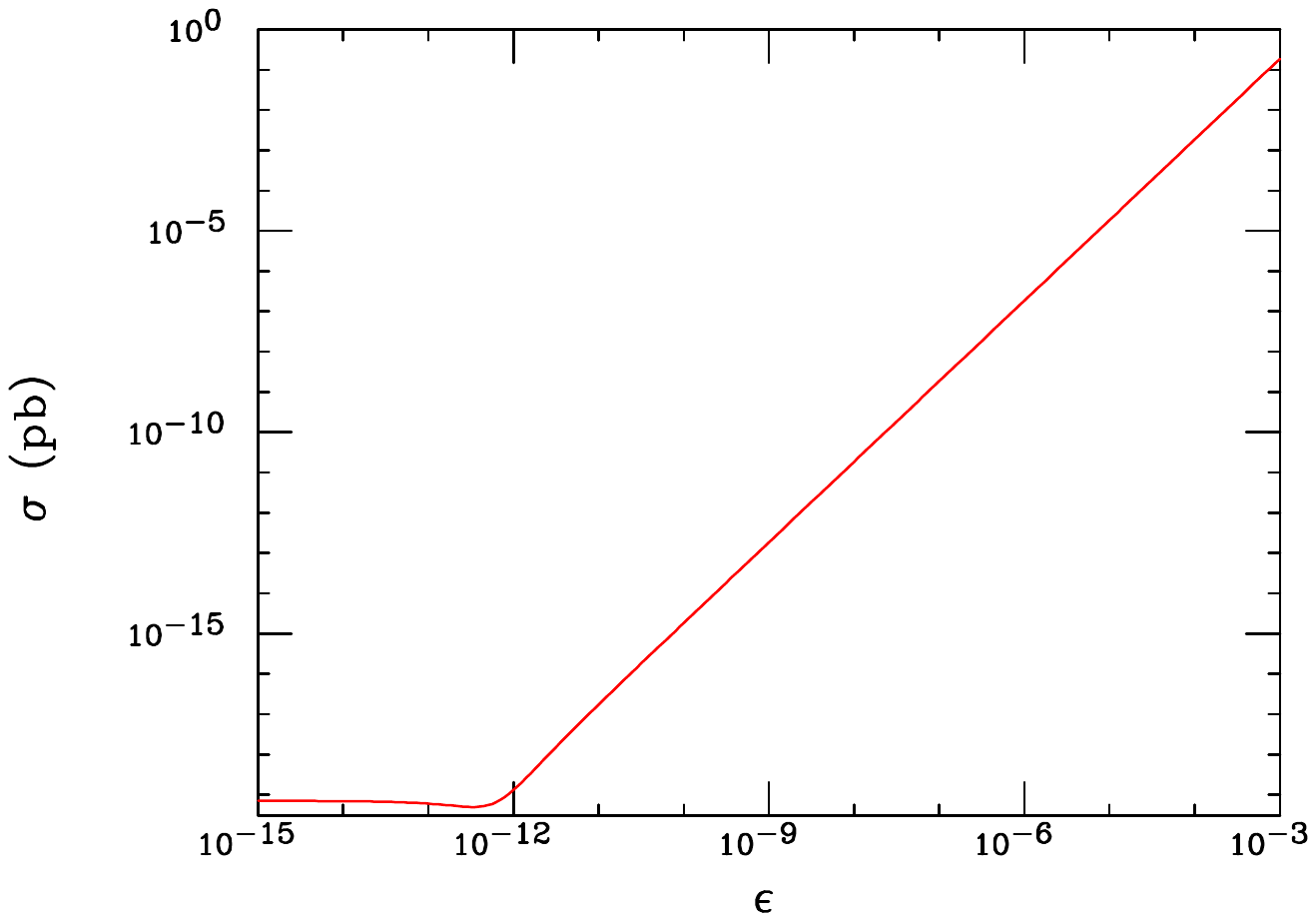}}
\vspace*{-1.50cm}
\caption{DM-electon scattering cross section as a function of $\epsilon$ assuming that $m_V=100$ MeV, $\beta_\phi=10^{-3}$, $\Lambda_e/\alpha_D=25$ TeV and $\alpha_D=0.5$ 
for purposes of demonstration, as discussed in the text; ${\cal G}=1$ has been assumed here. }
\label{fig11}
\end{figure}

\subsection{Fixed Target Searches}

Another tool which is used to access the parameter space of the dark photon and/or DM in KM models is DP production in fixed target experiments for which there are now 
many proposals in various stages of development\cite{Agrawal:2021dbo}. Here, we will follow the setup of the LDMX experiment at SLAC\cite{Akesson:2018vlm} whereat an electron 
scatters off of a heavy target nucleus (in this case Tungsten) emitting an on-shell DP in the process which may decay to either DM (which seems likely given the relic density calculation 
above) or to light SM states, \eg, $e^+e^-$, depending upon its mass. Specifically, in addition to rate issues, we are particularly interested in how the kinematic distributions for the DP (or, 
more precisely those of the visible recoiling electron) arising from an electron's dark magnetic dipole moment interaction differ from those obtained in the conventional KM setup with 
vector couplings. To be more specific, and in order to simplify our calculations, we will consider the (approximately forward) production of dark photons by the scattering an $E_0= 4$ GeV 
$e^-$ beam off of the Tungsten target employing the Improved Weizsacker-Williams Approximation as discussed in detail in Ref~\cite{Bjorken:2009mm} 
and compare the resulting distributions of these two extreme cases, \ie, the dark magnetic dipole interaction with that given by the familiar KM $\epsilon$ vector coupling term employing 
identical approximations in both cases.  Here, we will specifically follow the analysis and notation employed and presented in Appendix A of Ref~\cite{Bjorken:2009mm}. 

\begin{figure}[htbp]
\centerline{\includegraphics[width=5.0in,angle=0]{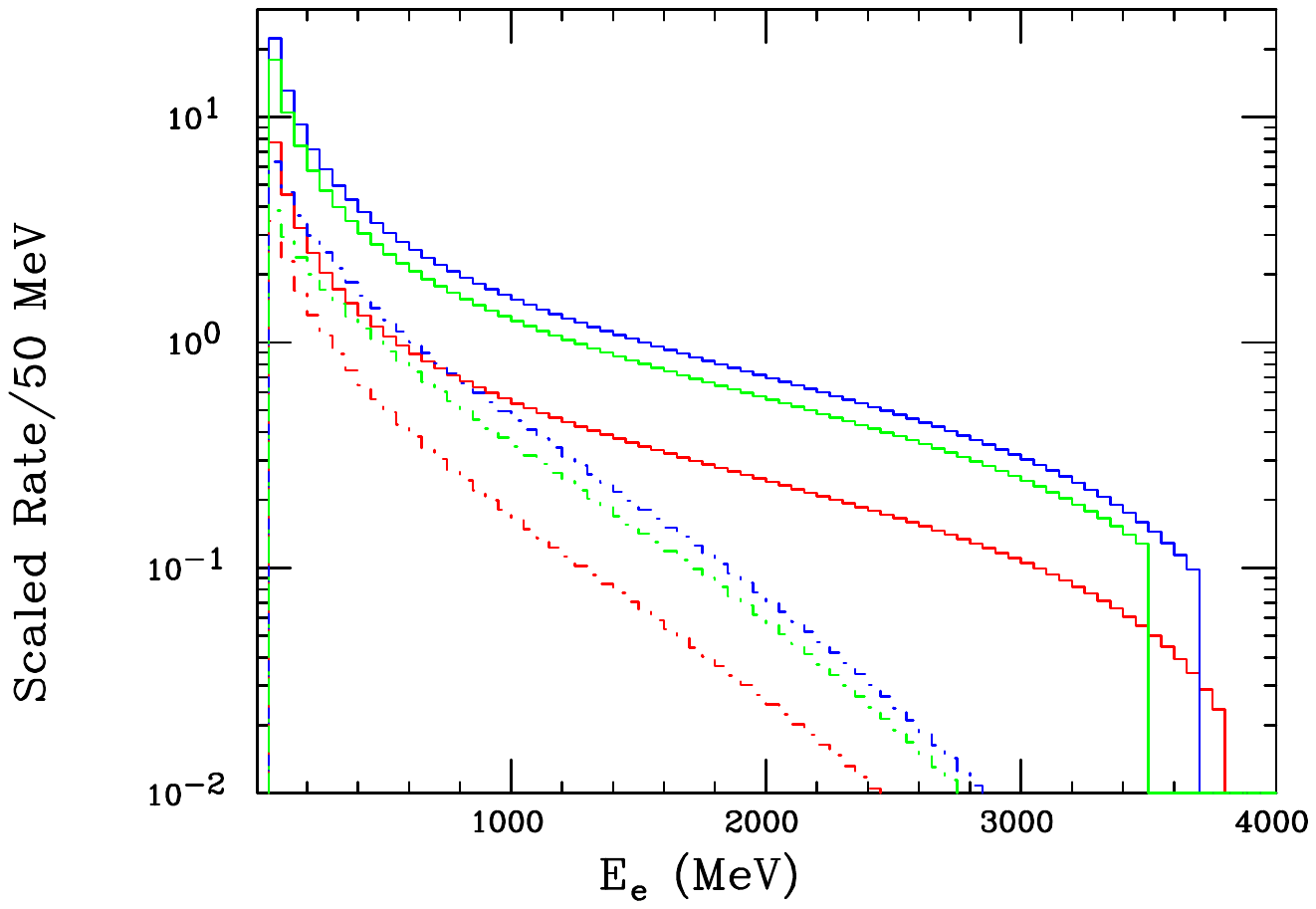}}
\vspace*{-2.3cm}
\centerline{\includegraphics[width=5.0in,angle=0]{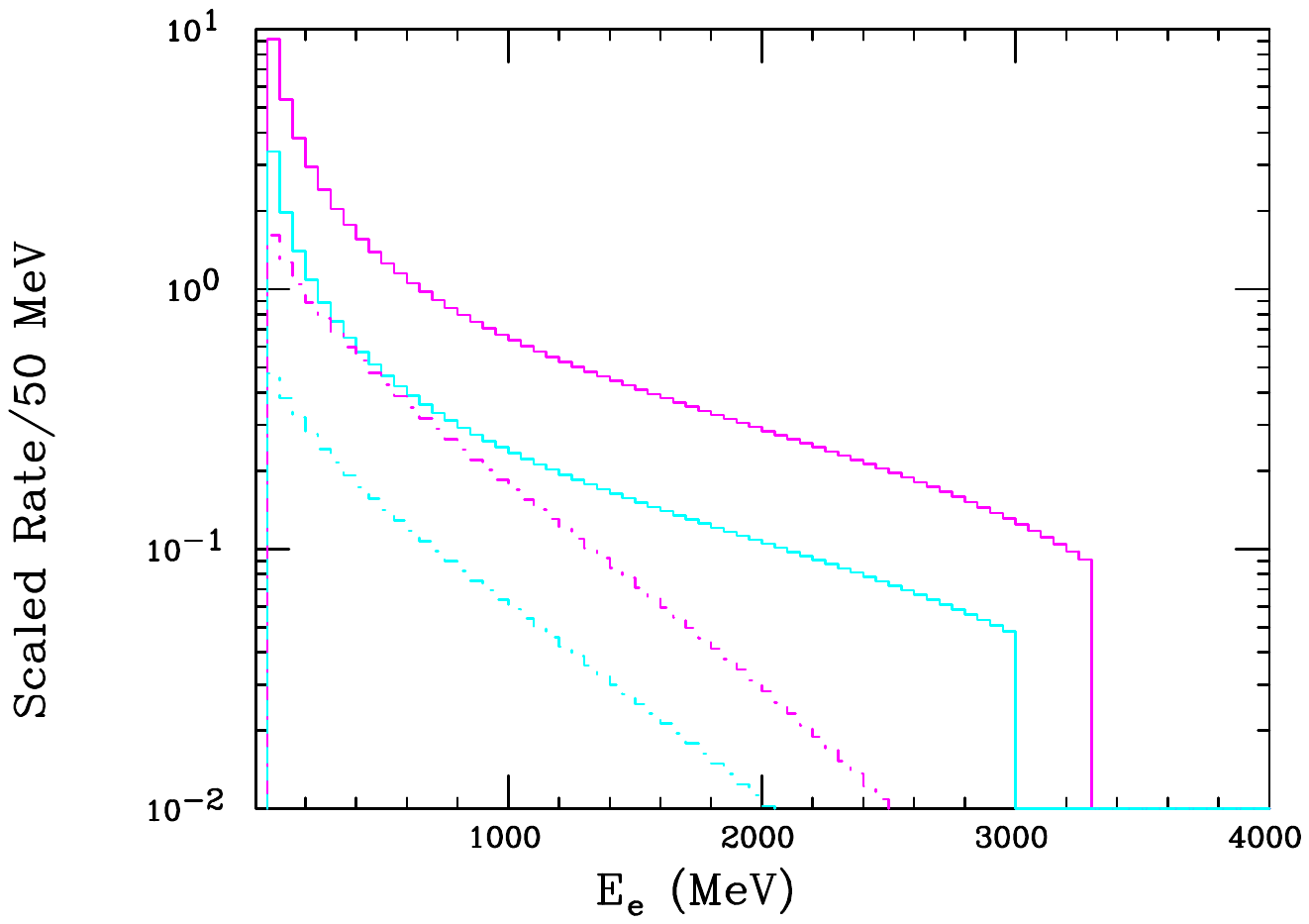}}
\vspace*{-1.30cm}
\caption{Comparison of the scaled electron recoil energy distribution for the familiar KM $\epsilon^2$ vector coupling scenario (solid) and the dark magnetic dipole model presently 
under consideration (dash dot)  assuming that $m_V=100$(red), 300(blue) or 500(green) MeV in the top panel and $m_V=700$(magenta) or 1000(cyan) MeV in the bottom panel, respectively. 
The normalization of all curves have been floated to fit these results onto these two panel to show their details and so only overall shape comparisons are relevant here.}
\label{fig4}
\end{figure}

In the initial step of the calculation, we determine the differential sub-process, $\gamma(q)e(p)\to V(k)e(p')$,  cross section for DP production in analogy and in comparison to 
Eq.(A11) of Ref.~\cite{Bjorken:2009mm}, to be (again, for the moment neglecting the mass of the electron) 
\begin{equation}
\frac{d\sigma}{d(p\cdot k)} = \frac{8\pi \alpha}{\alpha_D} ~\Big(\frac{\alpha_D}{\Lambda_e}\Big)^2~\Big(\frac{\hat s+\hat u}{\hat s^2}\Big)\,,
\end{equation}
with $-\hat u=U=2p\cdot k-m_V^2$ as given by their Eqs.(A7) and (A8) while $\hat s=2p'\cdot k+m_V^2$ is as given by their Eq.(A9). This directly leads to the double-differential 
production cross section for DP production, in analogy with their Eq.(A12), as
\begin{equation}
\frac{1}{E_0^2x} \frac{d\sigma}{dx d\cos \theta_V} = 8\alpha\tau \beta_E ~\frac{\alpha}{\alpha_D} ~\Big(\frac{\alpha_D}{\Lambda_e}\Big)^2~\Big[\frac{x}{U}\Big]\,,
\end{equation}
where $\theta_V$ is the angle between the incoming electron beam and the radiated DP in the lab frame, $\tau$ is a nuclear form factor\cite{Bjorken:2009mm}, 
$\beta_E^2=1-(m_V/E_0)^2$ and $x=E_V/E_0$ with $E_V$ being the DP recoil energy. As noted, in the small angle approximation, one obtains that 
$U=E_0^2\theta_V^2x+m_V^2(1-x)/x+m_e^2x$. Using the sample numerical input values as employed above, apart from pure kinematic effects which we will return to later, we 
can take the ratio of the pre-factors that appear so as to compare the overall DP emission strength arising from the dark magnetic dipole operator to that obtained from conventional 
$\epsilon^2$ suppressed vector coupling arising term from KM, obtaining:
\begin{equation}
{\cal R'}=\frac{1}{\alpha \alpha_D} \Big[\frac{\alpha_D m_V}{\epsilon \Lambda_e}\Big]^2\simeq \frac{0.22}{\alpha_D} \,,
\end{equation}
where we have imported the suggestive parameter values from before and which is unsurprisingly quite similar to the cross section ratio obtained above in the case of the relic 
density calculation. 

Given the similar production rates in these two models, we next need to compare the kinematic distributions of the scattered (visible) electron arising from the results above with 
those obtained from Eq.(A12) in Ref~\cite{Bjorken:2009mm}. Here, we are now only interested in the {\it shapes} of these distributions as their overall normalizations are reasonably 
similar as we have just determined. Fig.~\ref{fig4} compares the shapes of the various electron recoil energy distributions for the usual KM-induced vector coupling to that for the 
dark magnetic dipole moment interaction for dark photons with masses of $m_V=100,300,500,700$ and 1000 MeV, respectively. 
There are several things to observe here:  Since the dark magnetic dipole interaction coupling is 
directly proportional to the DP 4-momentum, one might imagine that the produced DP carries away more energy (and at larger angles) for this case than in the usual scenario of the 
vector KM interaction thus leading to a softening the electron recoil spectrum. Indeed, this is what we observe in this Figure in both of the top and bottom panels. Here we see that for all 
values of $m_V$ in our range of interest, while the overall spectrum shapes are approximately $m_V$-independent for both types of couplings, the electron recoil spectrum in the 
dark magnetic dipole case falls off much more swiftly with increasing $E_e$ than in the case where the ordinary KM-induced vector coupling of the DP is dominant. Clearly if {\it both} 
contributions were present we would expect a distribution of some intermediate shape. 

\begin{figure}[htbp]
\centerline{\includegraphics[width=5.0in,angle=0]{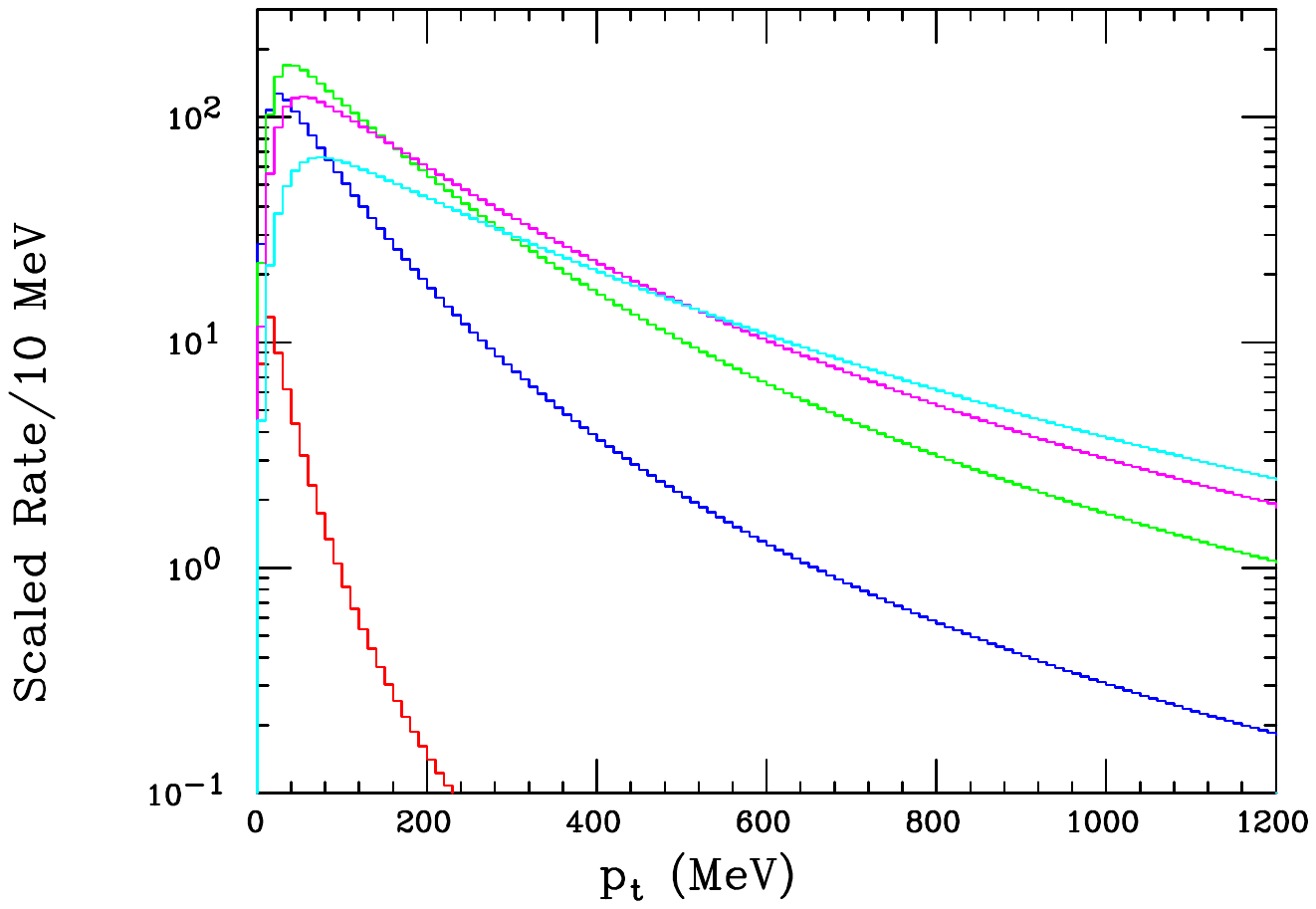}}
\vspace*{-2.3cm}
\centerline{\includegraphics[width=5.0in,angle=0]{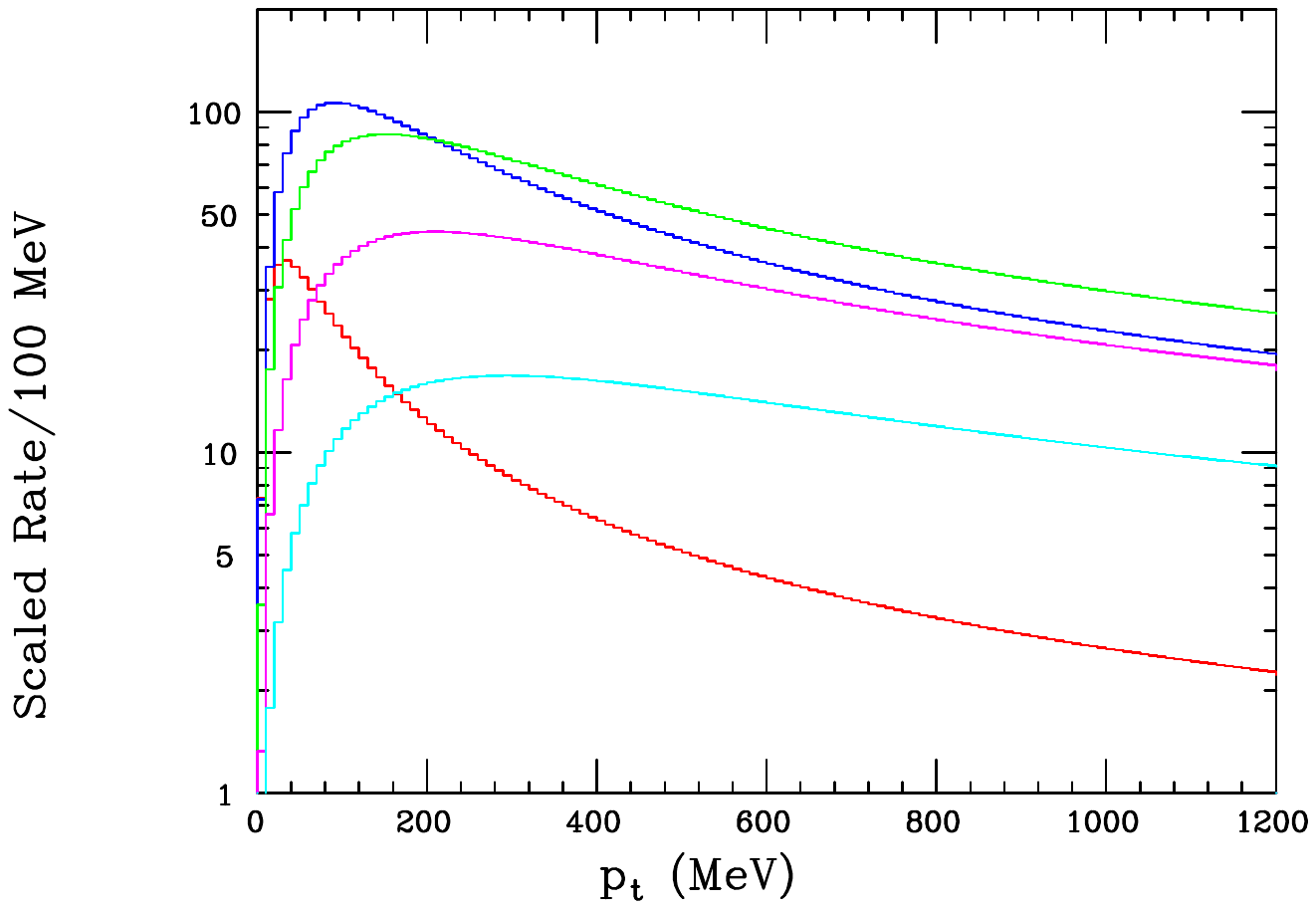}}
\vspace*{-1.30cm}
\caption{Comparison of the scaled recoil electron transverse momentum distribution for the (Top) familiar vector coupling in the KM/$\epsilon^2$ scenario and that from the dark 
magnetic dipole model presently 
under consideration (Bottom)  assuming that $m_V=100$(red), 300(blue), 500(green), 700(magenta) and 1000(cyan) MeV, respectively. The normalization of all curves have again been 
floated to fit these curves into single panels and to show their details as in the previous Figure so that only overall shape comparisons are relevant here.}
\label{fig5}
\end{figure}

A second similar distribution to examine is that of the recoil electron's transverse momentum, $p_T$, comparing those for the dark magnetic dipole moment with that arising from the 
usual KM-induced vector coupling. Due to the momentum dependence of the DP coupling in this case we now expect that its $p_T$ to be enhanced for fixed $m_V$ 
and thus, correspondingly due the forward production approximation made here, that of 
the recoiling electron is also identically enhanced. Fig.~\ref{fig5} compares the shapes of the electron $p_T$ distributions for the usual vector coupling (top) to that for the dark 
dipole moment interaction (bottom) for dark photons with masses $m_V=100,300,500,700$ and 1000 MeV, respectively. Again, there are several things to observe: \eg, in the KM case, 
the distribution maxima all lie at quite low values and then fall off very rapidly with increasing $p_T$, especially so at the lower end of the DP mass range that we consider here. In 
contrast to this, in the dark magnetic dipole case, the location of the distribution maximum is seen to increase with the value of $m_V$ and in all cases the distributions are seen to 
have significantly slower fall off, with much longer tails leading out to larger values of $p_T$.

While it appears that both classes of distributions are quite different and seem to allow for easy distinguishability at the level of this analysis for this range of DP masses, it is beyond 
the scope of 
the present work to determine in detail how sensitive experiments such as LDMX could be to these differences to separate these two distinct production processes.  This likely 
will require a much more sophisticated study including detector simulation. In particular, it may be of some interest to know if fixed target experiments can also determine the 
relative fractions of any admixtures of these two types of couplings if both were present simultaneously as in the cases described earlier above.

\subsection{Dark Photons from Meson Decays}

Meson decays with a DP in the final state, \eg, $\pi \to V\gamma$, offer another way to probe the parameter space of DP models. In practice, this turns out to be the most subtle of the 
constraints on the dark magnetic dipole model couplings that we will consider and requires some targeted work beyond what will be considered here within the context of a realistic 
PM model having properties similar to those above. In the usual KM scenario where the DP couples (in an $\epsilon$-suppressed manner) 
to the electromagnetic current, the relative branching fraction for this process is essentially just given by 
$B = \Gamma(\pi^0\to V\gamma)/\Gamma(\pi^0 \to 2\gamma) \simeq 2\epsilon^2 \tilde \beta^3$, where $\tilde \beta =1-m_V^2/m_\pi^2$ and follows 
immediately from the usual anomaly calculation. Similar results are obtained in the cases of, \eg, $\eta$ and $\eta'$ decays to the DP. In the dark magnetic dipole scenario, however, the DP 
no longer couples to the quarks in a simple vector-like manner to the electromagnetic current at the EFT level so that the usual scaling arguments following from the anomaly and leading to 
the straightforward branching 
fraction expression above are no longer applicable. These types of decays are actually quite different than those processes that we've so far encountered: since this dark magnetic 
moment coupling itself arises from a set of 1-loop diagrams similar to those in the toy model above, a complete calculation of the $\pi^0(\eta,\eta') \to V\gamma$ partial width/branching 
fraction forces us to consider an analogous set of 2-loop graphs together with the subtleties of the anomaly in a more realistic model, an analysis 
which is beyond the scope of the current paper. However, since either the DP mass, $m_V$, the pion decay constant, $f_\pi$, or the pion mass, $m_\pi$, all of which are of comparable 
magnitude, set the overall mass scale for this decay, one might make a {\it very} rough estimate for this branching fraction by naively assuming that the magnetic dipole interaction 
still remains dominant in an off-shell environment while 
at the same time being mindful that all the operators in Eq.(2), as well as several others, may now also play a potential role in generating the familiar $\pi^0$ decay amplitude.

To examine this idea further, we note that for off-shell fermions, assuming both P- and CP-conservation as we have above, the $\bar ffV$ vertex will in general consist of 12 distinct 
tensor structures\cite{Bincer:1959tz,Drell:1965hg,Naus:1989em}, each with its own form factor which at the 1-loop level will be a function of the three invariants 
$q^2,p^2$ and $p\cdot q$, in the notation of Eq.(1). This general vertex can be written in various convenient bases, \eg, using the product structure 
\begin{equation}
\Gamma^{\rm{off}}_\mu=\sum_{i,j}c_{ij}[q^2,p^2,p\cdot q]  ~(\gamma_\mu, p_\mu, q_\mu)_i \times (1, \qslash,\pslash,\qslash \pslash)_j\,,
\end{equation}
where the usual magnetic dipole moment operator can be obtained directly from the $c_{12}$ term via the well-known identity $\gamma_\mu \gamma_\nu=g_{\mu\nu}-i\sigma_{\mu\nu}$. The 
number of independent operators appearing here 
can be reduced to `only' 8 via the requirement of $U(1)_D$ gauge invariance and the application of the Ward Identities\cite{Bincer:1959tz,Drell:1965hg,Naus:1989em}. Subject to 
these same requirements, the toy model above is found to lead to an even smaller subset of just 6 of these 8, in principle, allowed tensor structures since we find that $c_{24,34}=0$. 
Given this, one reason we might speculate that the dark magnetic dipole operator may still dominate, at least within the framework of the toy model above, is that it is 
naively dimension-5 so that the corresponding form factor has an overall scaling of $m_F^{-1}$ while the other operators potentially contributing to the anomaly-type triangle graphs 
will generally instead scale as $m_F^{-2}$ as they are naively dimension-6. Some of the remaining operators may also produce the `wrong' gamma matrix and/or tensor structure to 
contribute to the hadronic matrix element for this decay process. 

Of course we cannot simply embed the magnetic dipole operator with a {\it constant} coefficient into the the $\pi^0$ decay graph without addressing the actual UV completion 
since it leads to a divergent result. Furthermore, various UV completions leading so the same effective value of this operator coefficient in the $q^2\, p^2,p\cdot q \to 0$ IR limit can 
produce different results in the case of $\pi^0$ decay\cite{Hambye:2021xvd}. Thus we need to have a realistic UV model in hand if we wish to make any detailed calculations. In any case, 
once we embed the full 1-loop off-shell vertex into the $\pi^0$ decay triangle graphs, the full loop momentum dependence of these form factors (arising from 
the original 1-loop graphs) will render their 2-loop contributions to these types of 
decays finite. Of course a detailed study is needed to elucidate these various expectations more fully. Assuming this dark magnetic dipole dominance to hold, in comparison to the result 
obtained above in the usual KM scenario for the $V\gamma$ branching fraction,  we can then estimate the relative size of these two contribution to this decay, ${\cal R_\pi}$, 
by employing dimensional analysis based on the results we have obtained previously above. Again quite naively, we might then expect this ratio to (very) roughly scale as 
\begin{equation}
{\cal R_\pi}\sim \frac{1}{\alpha \alpha_D} \Big[\frac{\alpha_D \cdot (f_\pi, m_V, {\rm or} ~m_\pi)}{\epsilon \Lambda_e}\Big]^2\sim \frac{0.1-0.2}{\alpha_D}\,,
\end{equation}
up to the additional O(1) corrections from the form factors mentioned above, after importing the suggestive numerical values as employed previously.  Thus we see, if this simple 
estimate is even approximately correct, that we might 
expect the branching fraction for the $V\gamma$ final state to be roughly the same as that in the usual KM, $\epsilon$-suppressed model with DM having only vector couplings. 
However, it is clear that a much more detailed estimate of this ratio should certainly be performed once a more realistic PM model becomes available.

\subsection{$e^+e^-$ Collider Searches}

$e^+e^-$ collisions offer a further way to access the parameter space of dark sector models via the production of dark photons/matter, usually in association with a tagging photon, \ie, in 
the $e^+e^-\to \gamma V$ process{\footnote {Here we will again work in the limit where the electron mass is taken to zero, \ie, $m_e \to 0$}}. As we will see below, such processes, due to the 
larger energy scales involved, \ie, $\sqrt s \simeq 10$ GeV, in comparison to those above offer the greatest challenge to the dark magnetic dipole moment scenario. Experiments at 
$e^+e^-$ flavor factories, \eg, BaBar, Belle, KLOE-2 and BESIII\cite{Lees:2014xha,Lees:2017lec,Ablikim:2018bhf,Ablikim:2017aab,Anastasi:2018azp,Kou:2018nap}
have performed and will continue to perform searches for these general types of processes in both the visible and invisible DP decay modes. In the case of invisible decays of the DP which  
are of more interest to us here, the current best constraint arises from BaBar\cite{Lees:2017lec}; in the usual KM scenario this requires that $\epsilon <\epsilon_{max} \lsim 10^{-3}$ for DP 
masses roughly lying in the interesting range $m_V \sim 0.1-1$ GeV.

Since 2-body phase space controls the energy of the photon in the final state, we cannot use this as a probe of the different natures of the two DP couplings to electrons in this case; however 
the different angular distributions may be useful in this regard. 
In the conventional KM model with vector couplings as well as in the dark dipole model we are considering here, this process occurs via $t$- and $u$-channel electron exchange. In the former 
case, we remind the reader that the associated poles in these channels will push both the $V$ and $\gamma$ into the forward/backward scattering directions so that detector 
angular acceptance cuts will reduce the signal rate. To see this, taking $z=\cos \theta$, the angle of the outgoing photon with respect to the incoming $e^-$, we recall that in the 
standard KM/vector coupling case the cross section for this process in the center of mass frame is given by\cite{Battaglieri:2021rwp} 
\begin{equation}
\frac{d\sigma}{dz}=\frac{2\pi \alpha^2\epsilon^2}{s} ~ \frac{\eta_V(1+z^2)+2(\eta_V^{-1}-1)}{(1-z^2)}\,,
\end{equation}
where here we define $\eta_V=1-m_V^2/s \simeq 1$ given the typical masses that we are discussing for the DP so that the second term essentially vanishes and the familiar QED-like 
singularities (with $m_e=0$ taken here) are clearly seen when $z=\pm 1$. 
We imagine that in a given experiment the integration over (just) the angular-dependent part of this expression subject to the necessary detector cuts just will yield a value of the 
angular acceptance, $A_V(\eta_V\simeq 1)$. These same poles are, however, absent in the dark magnetic dipole moment model due to the differing, momentum-dependent coupling 
structure and the corresponding cross section is instead given by (again still assuming $m_e=0$) 
\begin{equation}
\frac{d\sigma}{dz}=\frac{2\pi \alpha}{\alpha_D}~\Big(\frac{\alpha_D}{\Lambda_e}\Big)^2 ~\eta_V^2\,,
\end{equation}
which we see corresponds to a completely flat angular distribution. In this case, integration of this flat distribution over the same experimental angular acceptance cuts will yield the 
value $A_m(\eta_V\simeq 1)$ for the acceptance. Then the ratio of these two cross sections, taking these cuts into account, is similar in form to the many expressions encountered above 
and is just given by
\begin{equation}
{\cal R}_{e^+e^-}= \frac{1}{\alpha \alpha_D} ~\Big[\frac{\alpha_D \sqrt s}{\epsilon \Lambda_e}\Big]^2~ \frac{A_m}{A_V}\,,
\end{equation}
as might be expected. Now we can see immediately why the $e^+e^-$ search results are so restrictive: taking $\sqrt s=10$ GeV with $m_V^2/s \to 0$ as well as the the angular 
acceptances and familiar experimental constraints on $\epsilon$ as given in Ref.\cite{Lees:2017lec} into account in the same $\eta_V\to 1$ limit, we find the lower bound on the dark magnetic 
dipole scale to be given by 
\begin{equation}
\frac{\Lambda_e}{\alpha_D} \gsim ~ \frac{101~{\rm TeV}}{\sqrt \alpha_D}~\Big(\frac{10^{-3}}{\epsilon_{max}}\Big)\,,
\end{equation}
so that ($i$) the BaBar bound, $\epsilon \lsim 10^{-3}$, is seen to require the dark magnetic dipole scale to be at least 4 times larger than what we have been employing so far as a default 
and ($ii$) that if further null DP search results 
by Belle II are obtained this constraint will rapidly strengthen. Fortunately, we saw earlier that very large values of $\Lambda_e/\alpha_D$ are still allowed by the most important 
model constraint, \ie, that from obtaining the observed DM relic density and are certainly consistent with both the direct detection and the fixed target searches.

\begin{figure}[htbp]
\centerline{\includegraphics[width=5.0in,angle=0]{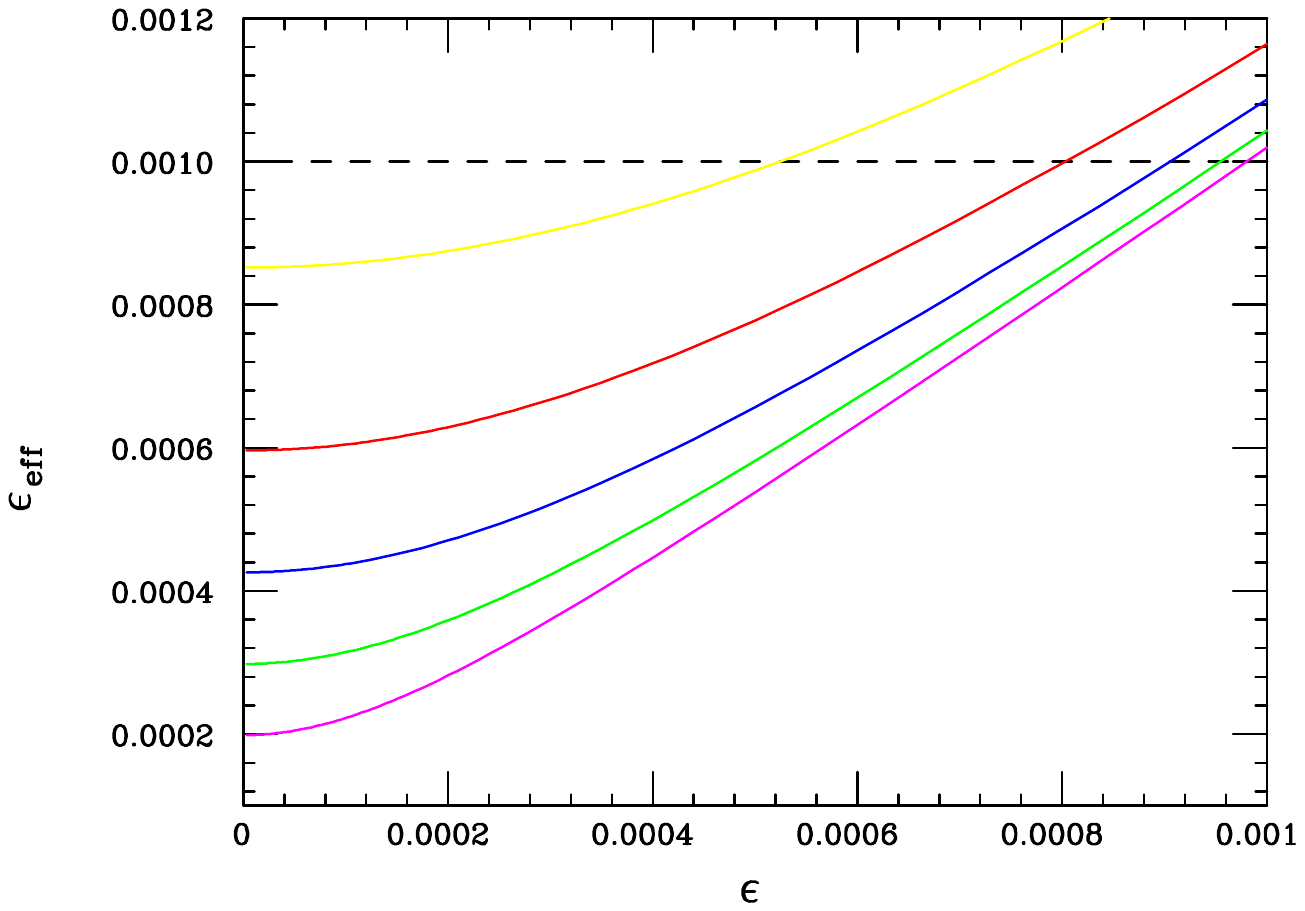}}
\vspace*{-1.50cm}
\caption{`Effective' value of $\epsilon$ obtained when both KM and dark magnetic dipole moments contribute to the $e^+e^-\to \gamma V$ process as a function of the true value of 
$\epsilon$ taking $\alpha_D=0.5$ for purposes of demonstration; the limit $m_V^2/s \to 0$ has been employed. The curves from top to bottom assume that 
$\Lambda_e/\alpha_D=175, 250, 350, 500$ and 750 TeV, respectively. 
The horizontal dashed line is the approximate current upper bound from BaBar for invisibly decaying DPs and so defines the presently allowed region of this parameter space. } 
\label{fig12}
\end{figure}

If restrictive detector acceptance cuts and/or significant SM backgrounds result in the reduction of the capability of future measurements to distinguish the potential KM and dark 
magnetic dipole moment contributions to this process from their angular distributions, any non-zero rate that is observed might be first interpreted as a finite value of 
an `effective' $\epsilon$ parameter, $\epsilon_{eff}$.  This may especially be true if both types of coupling structures 
were to be present simultaneously. To get an idea of how measurements would then impact this parameter space if such a possibility were realized, Fig.~\ref{fig12} shows the value of 
this $\epsilon_{eff}$ as a function of the true $\epsilon$ KM parameter for different assumed values of $\Lambda_e/\alpha_D$ in comparison to the present BaBar cross section 
constraint (here taking $\alpha_D=0.5$ for demonstration purposes) assuming the same angular acceptances from BaBar as noted above. When $\epsilon=0$ we recover the limit from the 
equation above whereas as $\Lambda_e/\alpha_D \to \infty$ we rather rapidly obtain the well-known KM model result.

As an important  reminder, we note that in this calculation above we have treated $\Lambda_e$ as a constant as we generally have in the relic density, fixed target and direct 
detection calculations discussed earlier. However, formally, in the present case of the $e^+e^-\to \gamma V$ 
process, one of the internal electron legs can go far off-shell, \ie, $|p^2_{\rm{internal}}| >> m_e^2$. Unlike in the discussion of $\pi^0$ decay, where the loop momenta can become very large 
and probe the underlying PM interactions themselves, here the maximum size of $p^2_{\rm{internal}}$ is controlled by the $e^+e^-$ collider energy which for BaBar and Belle remains 
rather small, \eg,  $\sqrt s \sim 10$ GeV, which is far below, \eg, the mass scales associated with the PM fields generating this interaction and thus we expect that this approximation 
of constant $\Lambda_e$ remains valid. However, as the collision energy increases this constant form factor approximation will break down: first, via the $p^2_{\rm{internal}}$-dependence 
of this form factor 
and then via the resolution of the PM loops themselves. Since we expect that $m_F \gsim 1$ TeV, once $\sqrt s$ approaches the energy range of the LHC or that of the possible future 
$e^+e^-$ colliders ILC, FCC-ee, CEPC and CLIC, we can no longer employ the approach as followed here. Clearly, then, we cannot simply extrapolate the type of bound that we have 
obtained above at $\sqrt s \simeq 10$ GeV in a trivial way to these much higher energy machines as one straightforwardly does in the usual KM model.

\section{Discussion and Conclusions}

In the usual Kinetic Mixing portal model, loops of portal matter particles carrying both SM and dark $U(1)_D$ gauge charges generate a mixing between the dark photon and the 
SM hypercharge gauge boson via vacuum polarization-like diagrams. If the PM fields are fermions, $F$, as was assumed in our study, they must have vector-like couplings to the 
SM gauge bosons in order to avoid gauge anomalies and also to avoid obtaining their masses from the SM Higgs vev. However, if at least some of the SM fermions, $f$, while 
remaining neutral under $U(1)_D$ (which has a coupling constant $g_D=\sqrt {4\pi \alpha_D}$),  lie in common representations of an extended dark gauge group with any of these 
PM fields, the additional 
new heavy dark gauge fields from this extended group can generate new types of interactions between the SM and the dark photon via 1-loop diagrams similar to those as 
shown in Fig.~\ref{fig1}. Such new loop-induced interactions may or may not dominate over the usual KM-induced interaction which we have found to generally be suppressed by a 
small parameter, $\epsilon \sim 10^{-(3-4)}$, in the PM framework. At energy scales in the IR far below the masses of these PM fermions, 
$m_F$ (and for at least approximately on-shell SM fermions), we can decompose these new interactions employing the familiar moment/form factor structure as is shown above in Eqs.(1) 
and (2). Assuming CP-conservation and large values for the PM masses consistent with LHC searches, $m_F\gsim 1$ TeV,  we might expect from this that the leading new interaction 
term arising at energies $<< m_F$ is that corresponding to a dark magnetic dipole moment with a scale parameter, $\Lambda_f \sim m_F/\alpha_D$, which has been the subject of our study. 
Although we briefly consider how $\Lambda_f$ might arise in our toy model, this quantity is treated as a constant, completely phenomenological parameter 
in most of our analysis. However, one must remain mindful that as the energy scale relevant to the processes under consideration, at either the tree-level or in loops,  grows larger 
this approximation will break down as scales $\sim m_F$ are approached and we'd then need to deal instead with the UV physics of the full underlying PM model. 

Within this framework, assuming the existence of a light DP as in the usual KM setup and a similarly light scalar DM field to easily satisfy the cross section bounds from the CMB via $p$-wave 
annihilation, we first determined the range of values of $\Lambda_e/\alpha_D$ as a function of $r=2m_{DM}/m_V$ for which the observed relic density can be obtained under the 
assumption that {\it only} the dark dipole moment interaction is active. If this scale parameter had been required to be small, \eg, implying that the PM should already have been 
observed, then this setup would have been immediately excluded. Fortunately, this was not the case: we found, depending somewhat on the value of $\alpha_D$ due to the effects of the finite 
DP resonance width, that values of $\Lambda_e/\alpha_D$ in the range $\sim 10-1000$ TeV can reproduce the observed relic density but with the exact value being rather strongly 
$r$-dependent. There was some preference found for the range $r<1$ implying that the DP would then decay invisibly in this setup. These results are, however, to a very good approximation, 
also found to be essentially {\it independent} of the value of $m_V$ itself -- unlike in the familiar KM picture -- up to factors of order unity due to new SM final state thresholds and DP finite 
width effects. It was 
also shown that if both the familiar KM-generated $\epsilon$ suppressed couplings as well as those arising from the dark dipole moment were {\it simultaneously} present that they would 
yield qualitatively similar numerical contributions to the DM annihilation cross section for typical values of the input parameters as might be expected in the PM framework.  
For fixed values of $m_V$ and $r$,  
a unique curve in the $\epsilon$-$\Lambda_e/\alpha_D$ plane is obtained which reproduces the observed DM relic density for various assume values of $\alpha_D$.

Armed with this somewhat restricted parameter range, we next examined the dark dipole moment model predictions for several other standard benchmark processes, making 
comparisons along the way with the corresponding results obtained in the familiar KM setup as well as asking what would happen if both interactions made contributions. In the case of 
direct detection, the contribution to the elastic scattering cross section from the dark dipole interaction is found to be very highly suppressed by the momentum-dependence of the operator 
itself. We demonstrated that even a very small value of $\epsilon$, far below the usual PM expectations, would completely overwhelm the pure dark magnetic dipole contribution to this 
process which by itself leads to an invisibly small cross section far below the neutrino floor. If both contributions were to be present, the conventional KM contribution would quite rapidly 
dominate even for quite small values of $\epsilon \gsim 10^{-12}$. 

The fixed target production window offers another way to access DP couplings; in particular, as a possible means to differentiate between DP to SM coupling structures given 
comparable statistics in either case. For typical parameter choices the rate for this process is roughly the same in the two purely KM or purely dark magnetic dipole limits but the $p_T$ 
and recoil energy distributions for the final state electron are found to be quite distinctive due to the momentum-dependence of the dark magnetic dipole vertex. In the later case, the 
recoil electron energies are found to be somewhat softer but the corresponding $p_T$ distributions are 
harder due to wider angle DP emission and the necessary $p_T$ balance. In the case of meson, \eg, $\pi^0$ decays to DPs, the situation is somewhat difficult to analyze since the 
loop generating this decay, unlike in the familiar KM scenario,  probes the UV physics details of the underlying structure of the PM model which leads to the dark magnetic dipole 
interactions themselves so that a complete study requires analyzing $\pi^0 \to V\gamma$ at the 2-loop level in a UV realistic model. Consideration of these underlying model-dependent 
graphs are necessary to obtain both a unique as well as finite result for the dark magnetic dipole coupling contribution to this partial width. Dimensional analysis, however, indicates that the 
rates anticipated in both the KM and dark magnetic dipole scenarios are most likely comparable up to O(1) factors. Finally, we considered the process $e^+e^-\to V\gamma$ at 
BaBar/Belle II energies $\simeq 10$ GeV, which are somewhat larger than those encountered in the previously examined tree-level processes. Because of this and the 
momentum-dependence of the dark magnetic dipole operator itself, this process is found to be the most constraining among those considered forcing us into the parameter range 
$\Lambda_e/\alpha_D \gsim100$ TeV from already existing BaBar measurements assuming that the DP decays invisibly. Future data from Belle II was shown to be able to provide even 
stronger constraints assuming a null search for this process. Performing similar analyses at $\sim$ TeV scale $e^+e^-$ colliders will most like require a consideration of the underlying 
UV model so that the lower energy limits from Babar/Belle are not easily generalized. 

The possibility that dark photons may interact with SM fermions in ways beyond those of the well-studied scenarios associated with kinetic and/or mass mixing with the photon and 
$Z$, respectively, and arising from loops of PM and other new particles is intriguing. Our preliminary study has only just scratched the surface of such a scenario.

\vspace*{0.7cm}

{\it Note Added} The possibility of dark dipole interactions for electrons in a purely EFT framework has also recently been considered in Ref.\cite{Barducci:2021egn}.

\section*{Acknowledgements}
The author would like to particularly thank J.L. Hewett, D. Rueter and G. Wojcik for very valuable discussions related to this work. He would like to thank D. Rueter in particular for 
generating Fig. 1. This work was supported by the Department of Energy, Contract DE-AC02-76SF00515.



\end{document}